\documentclass[10pt,a4paper]{article}
\usepackage{aer}
\usepackage{aer-osborn}
\usepackage{graphicx}
\usepackage[bf,hang,margin=7truemm]{caption}[2007/04/09]
\usepackage{amsmath}
\usepackage{bm}
\usepackage[]{natbib}
\begin{document}
\def\addnum#1{$^#1$}
\title{A Paradigm Shift from Production Function to Production Copula: Statistical Description of Production Activity of Firms}
\author{\sc Hiroshi Iyetomi$^1$\footnote{Corresponding author: 
{\tt hiyetomi@sc.niigata-u.ac.jp.}}, 
Hideaki Aoyama$^2$, Yoshi Fujiwara$^3$,\\ \sc
 Yuichi Ikeda$^4$, and Wataru Souma$^5$
}

\maketitle
\vspace{-25pt}
\begin{center}
{\sl
\addnum1 
Department of Physics, Niigata University, Niigata 950-2181, Japan\\[3pt]
\addnum2
Department of Physics, Kyoto University, Kyoto 606-8501, Japan\\[3pt]
\addnum3
ATR Laboratories, Kyoto 619-0288, Japan\\[3pt]
\addnum4
Hitachi Research Laboratory, Hitachi Ltd., Hitachi 319-1221, Japan\\[3pt]
\addnum5
College of Science and Technology, Nihon University, Funabashi 274-8501, Japan\\[3pt]
}
\end{center}

\begin{abstract}
Heterogeneity of economic agents is emphasized in a new trend of macroeconomics. Accordingly the new emerging discipline requires one to replace the production function, one of key ideas in the conventional economics, by an alternative which can take an explicit account of distribution of firms' production activities. In this paper we propose a new idea referred to as production copula; a copula is an analytic means for modeling dependence among variables. Such a production copula predicts value added yielded by firms with given capital and labor in a probabilistic way. It is thereby in sharp contrast to the production function where the output of firms is completely deterministic. We demonstrate empirical construction of a production copula using financial data of listed firms in Japan. Analysis of the data shows that there are significant correlations among their capital, labor and value added and confirms that the values added are too widely scattered to be represented by a production function. We employ four models for the production copula, that is, trivariate versions of Frank, Gumbel and survival Clayton and non-exchangeable trivariate Gumbel; the last one works best.
\end{abstract}

\noindent \textit{Keywords}: Copula; Correlation; Production Function; Firm; Value Added

\centerline{JEL No: D24, D30, L11CE10}
\vspace{3pt}
\centerline{KUCP-2183}

\newpage
%%%%%%%%%%%%%%%%%%%%%%%%%%%%%%%%%%%%%%%%%%%%%%%%%%%%%%%%%%%%%%%
% local definitions.
\newcommand\mdef{:=}
\newcommand\mat[1]{\mathbf{#1}}  % matrix
\newcommand\T{{\scriptstyle\mathsf{T}}}  % transpose of matrix/vector
\newcommand{\avg}[2][]{\left\langle{#2}\right\rangle_{\scriptstyle{#1}}}
% leqslant is missing. the following is temporaly definition.
\newcommand\leqslant{-?-}
%%%%%%%%%%%%%%%%%%%%%%%%%%%%%%%%%%%%%%%%%%%%%%%%%%%%%%%%%%%%%%%
\section{Introduction}

Recently a new approach to macroeconomics has emerged~(\cite{AY2007}, \cite{DGGGP2008}, \cite{AFIIS2010}) by taking serious account of heterogeneity of economic agents. Various stylized facts have accumulated to support this new perspective, including the distribution of firms' size with a power-law tail. The traditional concepts in mainstream economics should thus be replaced by new ones: a representative agent, by heterogeneous agents; average, by distribution; deterministic, by stochastic; mechanical point of view, by statistical mechanical point of view; and so forth. The principal goal of the new approach is to provide microscopic foundations for macroeconomics. Statistical mechanics in physics is a discipline to understand macroscopic states of matter using microscopic information on atoms and molecules. It is a good successful model for the endeavor to bridge between microeconomics and macroeconomics.

In this paper we pay our attention to production activities of firms. The production function is one of basic ingredients in microeconomic theory~(\cite{Wicksteed1894}, \cite{Varian1992}). It specifies output $Y$ of a firm for given input factors such as labor $L$ and capital $K$:   
\begin{equation}
Y=F(L,K),
\label{yf}
\end{equation}
where it is assumed that each firm produces goods in an optimized manner with its own production function. On the other hand, an aggregated production function is also an important tool to measure production activities at national level in macroeconomic theory~(\cite{sato1975}). The gross domestic product (GDP), given as the total sum of value added of producers over nation, is assumed to be a function of aggregated capital and labor for instance. An aggregated production function is in principle obtained by summing up all the production functions of individual producers. However this naive definition is an impractical way to construct the GDP; actually it is estimated through the system of national account (SNA). There are also methodological problems associated with the aggregation, including how to define aggregated capital and labor. Thus the production function at a macroscopic level has no sound microscopic foundation yet.

According to the spirit of the newly emerging economic discipline, a right direction to proceed is to reproduce observed data as they are. The analysis of financial data of listed firms in Japan shows that there are significant correlations among their capital, labor and value added; we regard those firms as constituting a statistical ensemble. Then we model such correlations in the production variables of firms using copulas~(\cite{sklar1959}). Copulas are analytic functions created in statistics to describe genuine correlations among variables~(\cite{Nelsen2006}, \cite{Joe1997}). Mathematical properties of copulas have been studied extensively, although the concept of copulas has only a few decades of active history. Recently copulas have also attracted interest from business practitioners who have in their mind a number of possible applications to finance such as asset allocation, credit evaluation, default risk modeling, derivative pricing, and risk management~(\cite{Cherubini2004}, \cite{MS2006}).

The copula model so obtained predicts value added yielded by a firm with given capital and labor in a probabilistic way. The new idea referred to as production copula is in sharp contrast to the production function which is completely deterministic. The present paper can thus be regarded as a generalization of the canonical economic concept at a microscopic level. And hence the production copula is expected to provide a tool to understand economic phenomena at both microscopic and macroscopic levels on an equal basis.

In the following section we explain financial data on Japanese listed firms which are used in this paper. In Sec. III we fit the data to the production function of Cobb-Douglas form and demonstrate the real data cannot be accommodated in the framework of the production function. In Sec. IV we prepare the statistical modeling of production activities of firms in terms of copulas by briefly reviewing the concept of copulas. Section V is devoted to actual construction of the production copula. In Sec. VI we conclude this paper by pointing out possible applications of the production copula obtained.

\section{Data on Listed Firms in Japan}
The present analysis of real data is based on the NEEDS
database~(\cite{nikkei}). It has accumulated financial statement data of listed firms in Japan over last 30 years; it is very exhaustive as regards financial information on the listed firms. We selected firms belonging to the manufacturing sector and compiled the fundamental quantities $L$, $K$, and $Y$ for the production activities of those firms. To have a concrete idea, we actually substituted $L$, $K$, and $Y$ respectively with labor cost, fixed asset, and value added; all of them are measured in units of million yen. Since labor cost $L$ and fixed asset $K$ are primary quantities, there is no ambiguity for them. However, there are basically two alternative ways (subtractive or additive method) to derive value added $Y$ leading to different results. We refer the readers to \cite{soumaikeda} for details about the calculation of $Y$. We use the data set in 2006 throughout this paper; the total number of firms is $N=1360$.\footnote{We have excluded one manufacturer with negative $Y$ out of our database.}

\begin{figure}[!tp]
\begin{center}
\includegraphics[width=0.5\textwidth,bb=0 0 1347 2541]{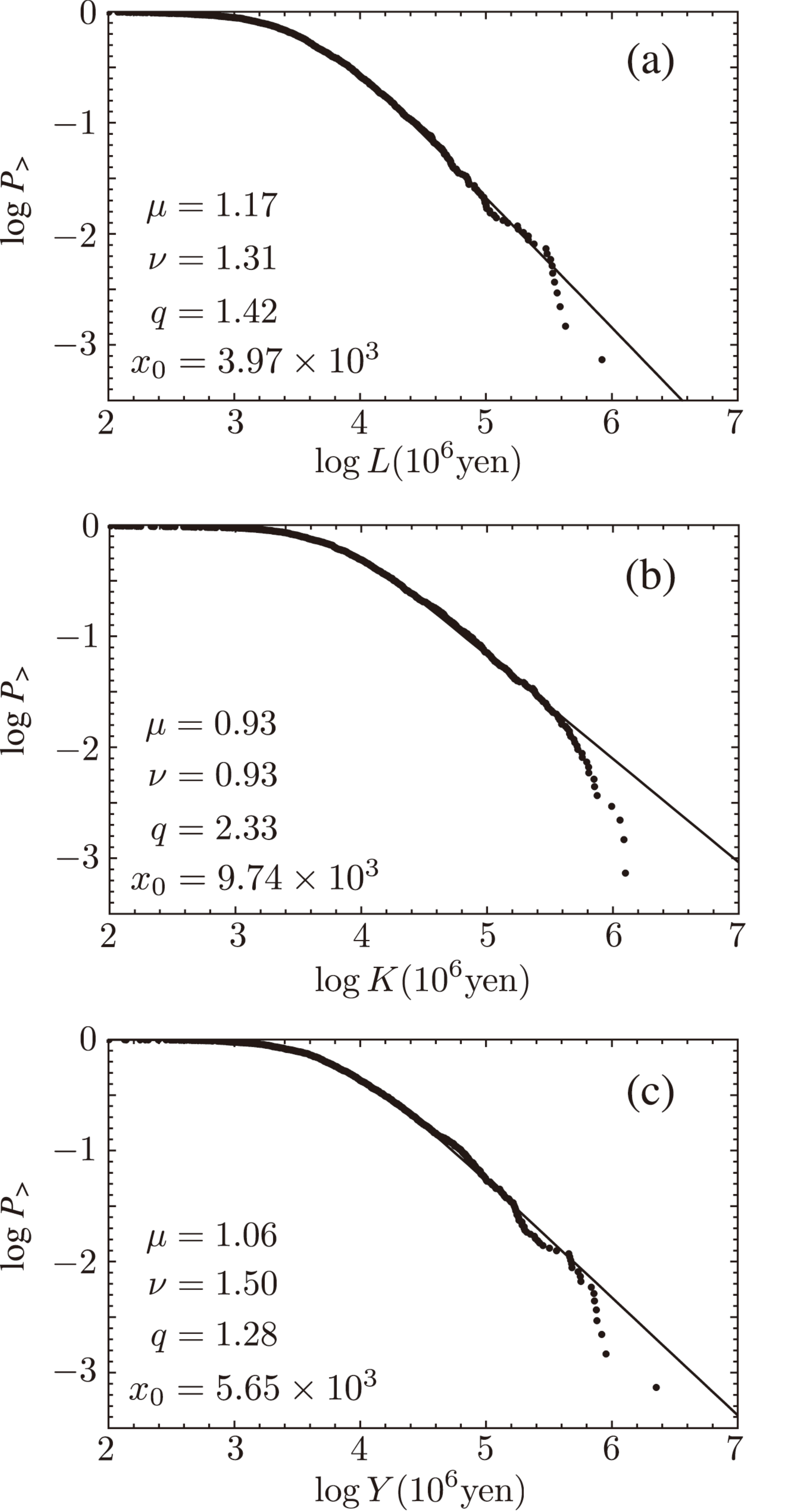}
\caption{Double-logarithmic plots of the complementary CDF's for the labor cost (a), the capital (b), and the value added (c) of the Japanese listed firms in the manufacturing sector, as depicted by dots. Results fitted to the real data in the form (\ref{bb2}) are also shown by a solid line in each panel, along with the model parameters determined by the MLM.}
\label{distKLY_NEEDS2006}
\end{center}
\end{figure}

In Fig.~\ref{distKLY_NEEDS2006} we plot the complementary cumulative distribution functions (CDF's) for $K$, $L$, and $Y$ of the manufacturers. These figures clarify that the distribution of each financial quantity has a power-law tail\footnote{To claim this fact, we discard the data points for the top 1\% of firms in each panel of Fig.~\ref{distKLY_NEEDS2006} which are considerably depressed as compared with the power-law behavior. Such a cutoff may be ascribed to finite size effects in data collection.} . We fitted these data adopting the following form, called the generalized beta distribution of the second kind (\cite{McDonald1984, KK2003}), for the probability density function (PDF) $p(x)$:
\begin{equation}
p(x; \mu, \nu, q, x_{0})=\frac{q}{B(\mu/q,\nu/q)} \frac{1}{x} \left( \frac{x}{x_0} \right)^{\nu}\left[\,1+\left( \frac{x}{x_0} \right)^q\right]^{-(\mu+\nu)/q},
\label{bb2}
\end{equation}
where $B(r,s)$ is the beta function, defined as
\begin{equation}
B(r,s):=\int_{0}^{1}u^{r-1}(1-u)^{s-1}du.
\label{betafunc}
\end{equation}
The fitting parameters $\mu, \nu, q$ and $x_{0}$ are assumed to take values in $\mu > 0, \nu > 0, q > 0$ and $x_{0} > 0$. The complementary CDF $P_>(x)$ corresponding to Eq. (\ref{bb2}) is 
expressed explicitly using the incomplete beta function $B(z,r,s)$ as
\begin{equation}
P_>(x) := \int_x^\infty {dx'} p({x'})=\frac{B(z,\mu/q,\nu/q)}{B(\mu/q,\nu/q)},
\label{bbdef}
\end{equation}
with
\begin{equation}
z=\left[\,1+\left( \frac{x}{x_0} \right)^q\right]^{-1},
\label{bbdef1}
\end{equation}
where
\begin{equation}
B(z,r,s):=\int_{0}^{z}u^{r-1}(1-u)^{s-1}du.
\label{incompletebetafunc}
\end{equation}
The parameters $\mu$ and $\nu$ are power-law exponents in the large and small $x$ limits, respectively:
\begin{equation}
p(x) \propto
\begin{cases}
\displaystyle \biggl( {\frac{x}{{x_0 }}} \biggl)^{ - \mu  - 1} & {\textrm{for}\ x \to \infty}\, , \\
\displaystyle \biggl( {\frac{{x_0 }}{x}} \biggl)^{ - \nu  + 1} & {\textrm{for}\ x \to 0}\, . \\
\end{cases}
\end{equation}
The parameter $x_0$ represents a characteristic scale and $q$ is a crossover parameter connecting the two limiting regions. Figure~\ref{distKLY_NEEDS2006} confirms that Eq.~(\ref{bb2}) works well to reproduce the original data including the fat tail, where we determined the parameters using the maximum likelihood method (MLM).

The likelihood is generally defined by
\begin{equation}
L \left( \theta_1, \cdots , \theta_{k} \right) = \prod_{i=1}^{n} p \left( x_{i}; \theta_1, \cdots , \theta_{k} \right) \, ,
\label{eq:likelihood}
\end{equation}
where $\{ x_{i} \}$ are $n$ observed values and $p \left( x; \theta_1, \cdots , \theta_{k} \right)$ is a model PDF with $k$ fitting parameters $\{ \theta_{j} \}$. It gives the joint probability density function for all the observations, being assumed to be independent and identically distributed. An optimum set of the parameters in the model PDF is then obtained through maximization of the likelihood. This illustrates the MLM for fitting a model PDF to given data, which we use throughout this paper. The log-likelihood defined by 

\begin{equation}
\ell \left( \theta_1, \cdots , \theta_{k} \right) := \log L \left( \theta_1, \cdots , \theta_{k} \right) = \sum_{i=1}^{n} \log p \left( x_{i}; \theta_1, \cdots , \theta_{k} \right) \, ,
\label{eq:log-likelihood}
\end{equation}
is usually adopted as an objective function to be maximized in place of the likelihood $L$.

\begin{figure}[!tp]
\begin{center}
\includegraphics[width=\textwidth,bb=38 382 581 575]{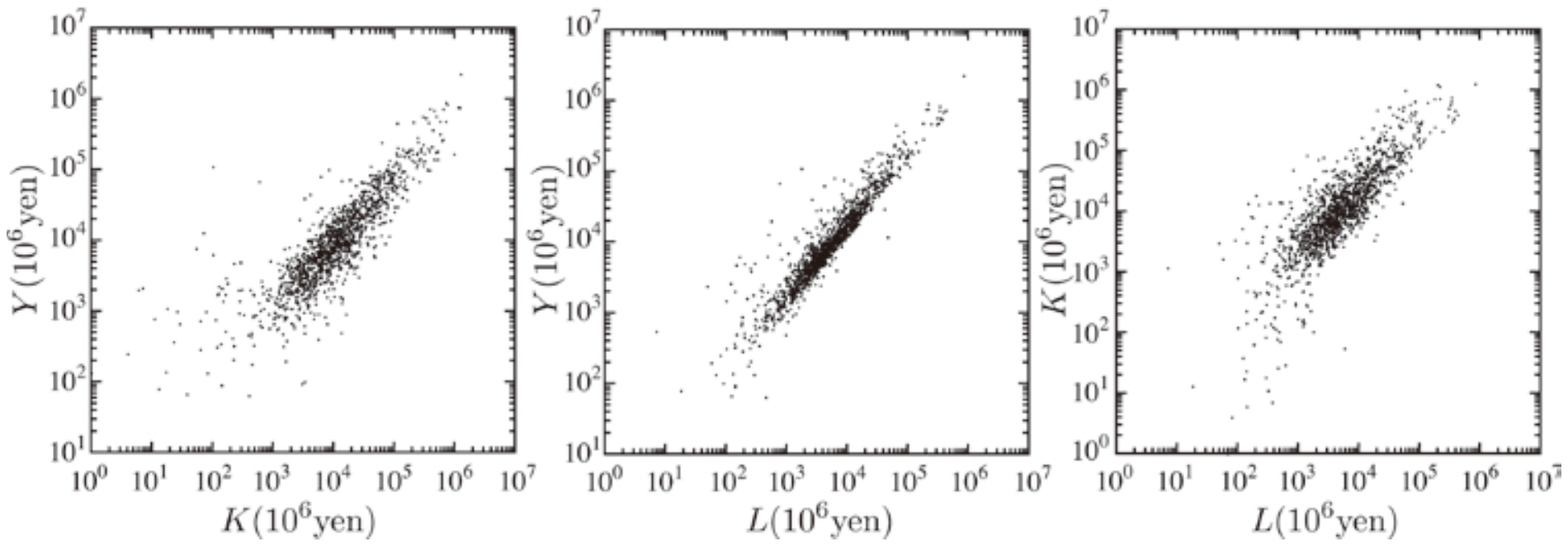}
\caption{Correlations for each pair of the three financial quantities, labor cost $L$, capital $K$, and value added $Y$, of the manufacturers in Japan. Note that these are double-logarithmic plots.}
\label{KLY2_NEEDS2006}
\end{center}
\end{figure}

In Fig.~\ref{KLY2_NEEDS2006} we show scatter plots for all pairs of the three financial quantities in the manufactures. We see that those variables are mutually correlated, assuming that those firms constitute a statistical ensemble. Although the inputs $L$ and $K$ are mathematically treated as independently controllable variables in the production function, those are statistically correlated to a certain extent. Such dependence between $K$ and $L$ is beyond the scope of the production function itself. A ridge theory of the production function elucidates that the dependence between the inputs arises from profit maximization behavior of firms (\cite{souma:nfa}, \cite{AFIIS2010}). Spearman's rank correlation coefficient $\rho_{\mathrm{S}}$ is one of non-parametric measures of correlation between two variables. The values of $\rho_{\mathrm{S}}$ for the three pairs, $K$-$Y$, $L$-$Y$, and $L$-$K$, are calculated as 0.86, 0.95, and 0.83, respectively. Thus the $L$-$Y$ pair has stronger rank correlation than the remaining two pairs $K$-$Y$ and $L$-$K$ have. And the remaining ones have rank correlations of similar strength. The detailed correlation structure for each pair is described in terms of copulas later, with an account of these results for the overall correlations.

\section{Failure of the Production Function}
Before moving to the main theme in this paper, we spend short time here to demonstrate how inappropriate is the concept of the 
production function to describing the data used here. One of the simplest functional forms for the production function (\ref{yf}) is given as
\begin{equation}
F(L,K)=AK^{\alpha}L^{\beta},
\label{eq:PF_CD}
\end{equation}
where $A$, $\alpha$, and $\beta$ are adjustable parameters to fit data. This functional form was introduced by \cite{CD1928} a long time ago and has been extensively adopted because of its ease of use and its extreme flexibility. In addition we note that a generalized form, referred to as the constant elasticity of substitution (CES) production function, is also available~(\cite{CES1961}).  

\begin{figure}[!tp]
\begin{center}
\includegraphics[width=0.9\textwidth,bb=0 0 1848 889]{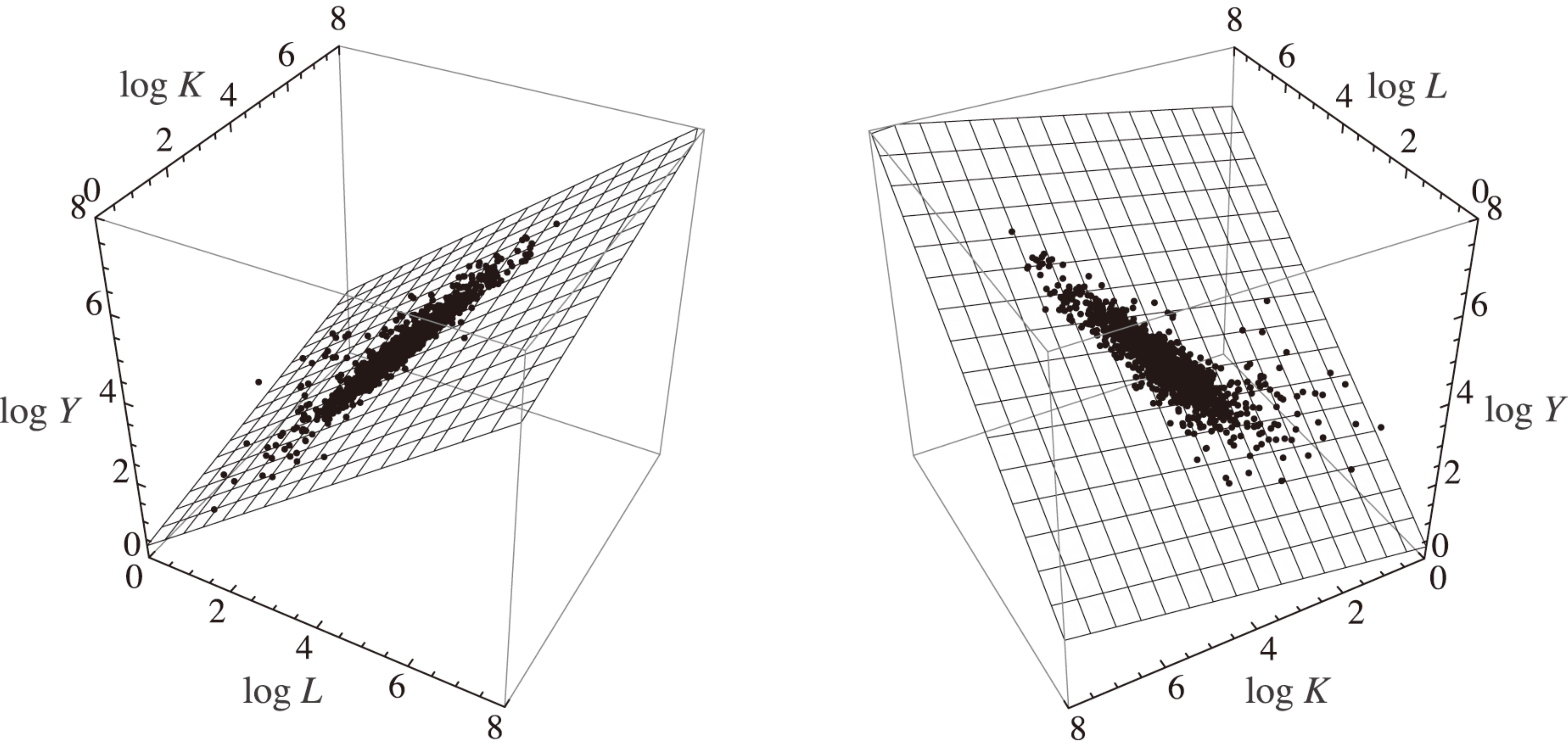}
\caption{The CD production function (mesh plane) fitted to the real data (dots) in $L$-$K$-$Y$ space viewed from two different angles. Note that all of the axes are in a logarithmic scale.}
\label{fig:pf_CD}
\end{center}
\end{figure}

We have fitted the Cobb-Douglas (CD) form (\ref{eq:PF_CD}) to the financial data of the listed firms using the least square method and have obtained the following values:
\begin{equation}
A=2.160,\quad \alpha=0.183, \quad \beta=0.788 \; .
\label{eq:noncons}
\end{equation}
We note that the production function almost satisfies the homogeneity condition of degree one, namely, exhibits constant returns to scale:
\begin{equation}
\alpha+\beta=0.971 \; .
\label{eq:alphabeta}
\end{equation}
The CD production function calculated with these parameters is compared with the original data in Fig.~\ref{fig:pf_CD}. Although one might be tempted to conclude that the CD
production function describes the data well in viewing Fig.~\ref{fig:pf_CD}, it is far from the truth as demonstrated below. Such a figure drawn with all of the axes in a logarithmic scale may be useful to have a bird's eye view, but is misleading in the
sense that huge differences of actual numbers are trivialized by
taking logarithm.

\begin{figure}
\begin{center}
\includegraphics[width=0.8\textwidth,bb=27 314 546 569]{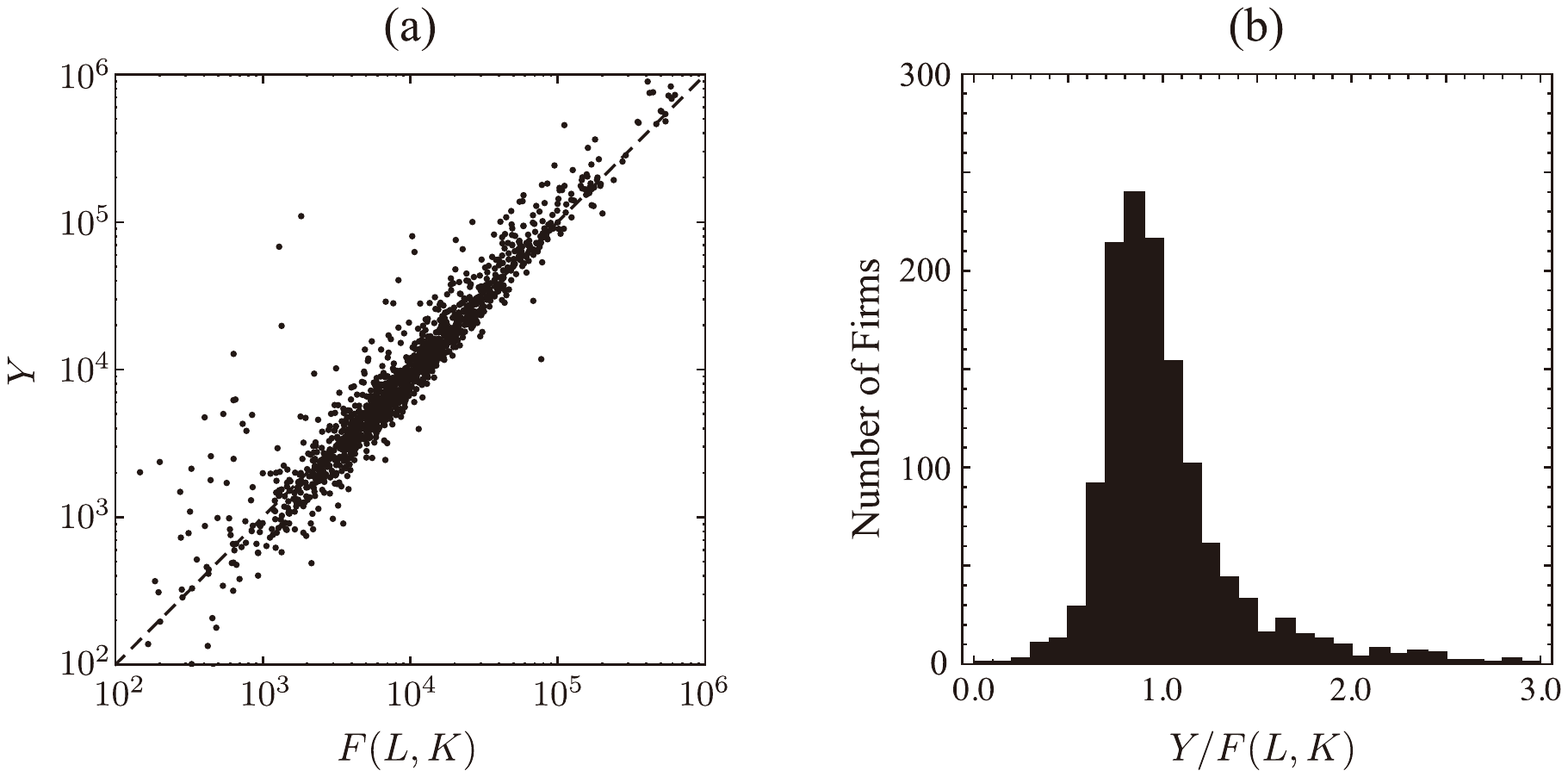}
\caption{Difference between the actual $Y$ and
the best-fit CD function $F(L,K)$:
(a) double-logarithmic scatter plot of the two variables, where the dashed line shows the diagonal line to gauge accuracy of the CD function; 
(b) histogram of their ratio with bin size of $0.1$.}
\label{fig:yycd}
\end{center}
\end{figure}

\begin{figure}
\begin{center}
\includegraphics[width=0.8\textwidth,bb=80 315 539 540]{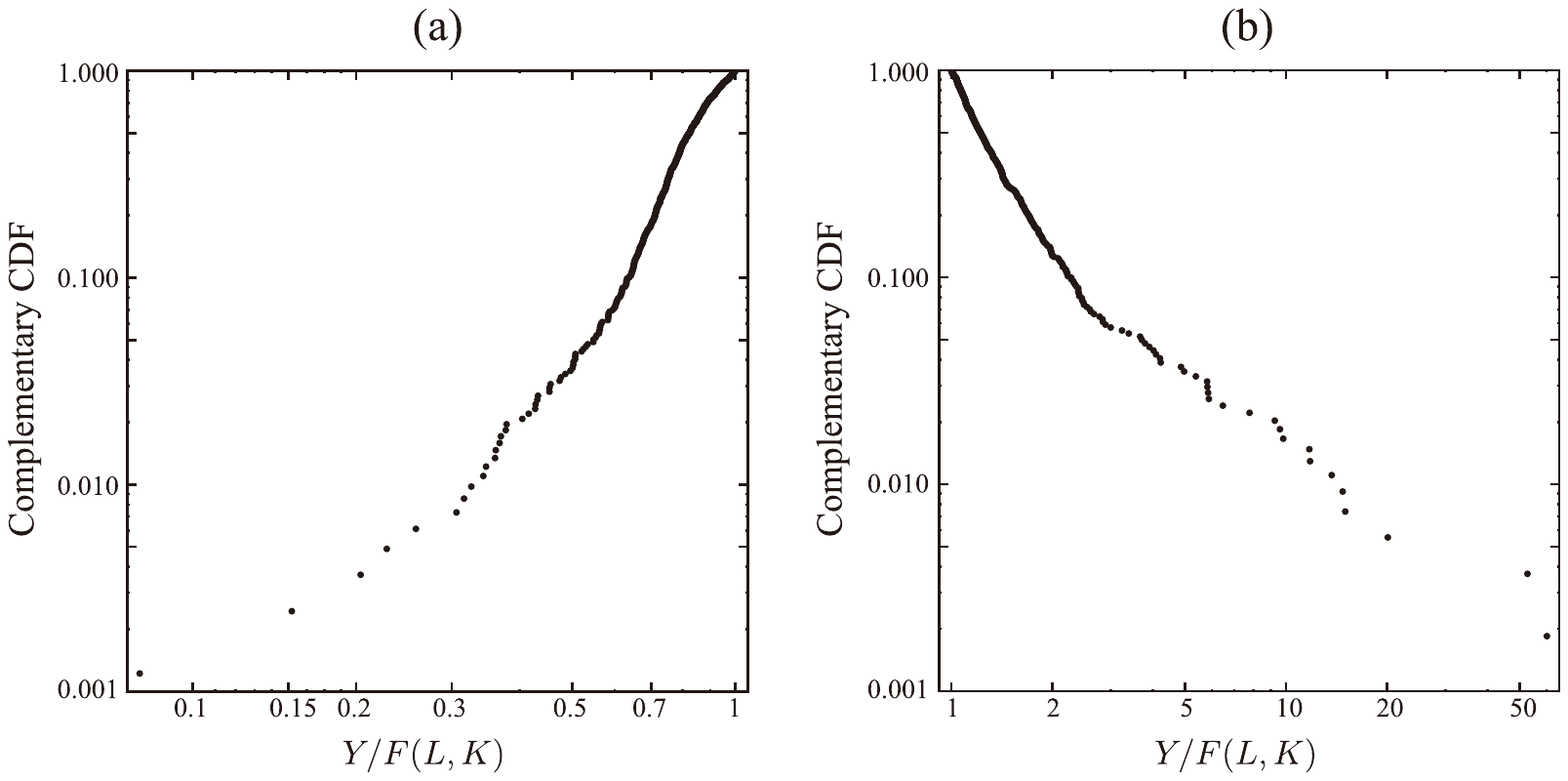}
\caption{Complementary CDF's of $Y/F(L,K)$
on the lower side $[0,1]$ (a) and the higher side $[1,\infty]$ (b), where the CD function is used. Note that both panels are double-logarithmic plots.}
\label{fig:yycdfat}
\end{center}
\end{figure}

Figure~\ref{fig:yycd}
shows difference between the original values of 
$Y$ and the corresponding results $F(L,K)$ obtained by the best-fit CD production function with the parameters as
given in Eq. (\ref{eq:alphabeta}). In the scatter plot (a) as well as in the histogram (b), we observe that the
actual data is widely scattered around $F(L,K)$. If the CD production function worked perfectly, the dots would be aligned along the diagonal line in Fig.~\ref{fig:yycd}(a) and the distribution of $Y/F(L,K)$ would be of a delta-function shape centered at $Y/F(L,K)=1$ in Fig.~\ref{fig:yycd}(b).

The wide discrepancies between the actual data and the predictions due to the CD production function are more clearly visible in Fig.~\ref{fig:yycdfat}, which plots the complementary CDF's of $Y/F(K,L)$ by separating its region into two sides with the boundary $Y/F(K,L) = 1$. Either of the CDF's is shown to have linear-like behavior in a double logarithmic plot, indicating that the distribution functions decay very slowly. Because of these fat tails, the production 
function often fails in reproducing the actual output $Y$ by large ratio. For instance, 41\%(27\%) of 540 firms have $Y$ values which are more than 30\%(50\%) larger than the corresponding values of $F(K,L)$ on the higher side ($Y/F(K,L) >1$) and 18\%(3.5\%) of 820 firms have $Y$ values more than 30\%(50\%) smaller than the $F(K,L)$ values on the lower side ($Y/F(K,L) <1$). While the production function describes the central values of the distribution, it by no means can be used for any quantitative theory involving $Y$.

\begin{figure}[!tp]
\begin{center}
\includegraphics[width=0.9\textwidth,bb=0 0 1848 881]{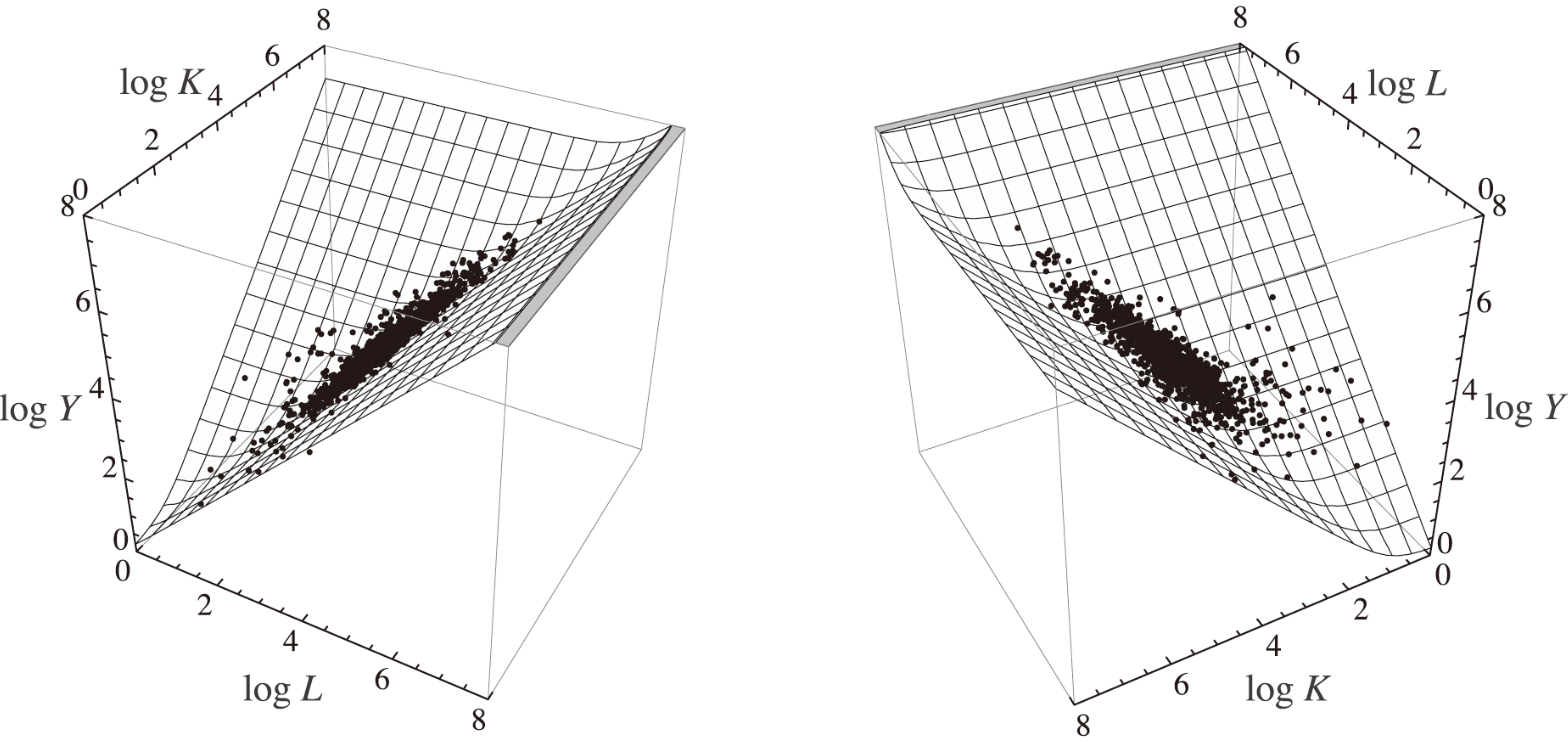}
\caption{The same as Fig.~\ref{fig:pf_CD}, but where the CES production function is used.}
\label{fig:pf_CES}
\end{center}
\end{figure}

\begin{figure}
\begin{center}
\includegraphics[width=0.8\textwidth,bb=47 293 544 537]{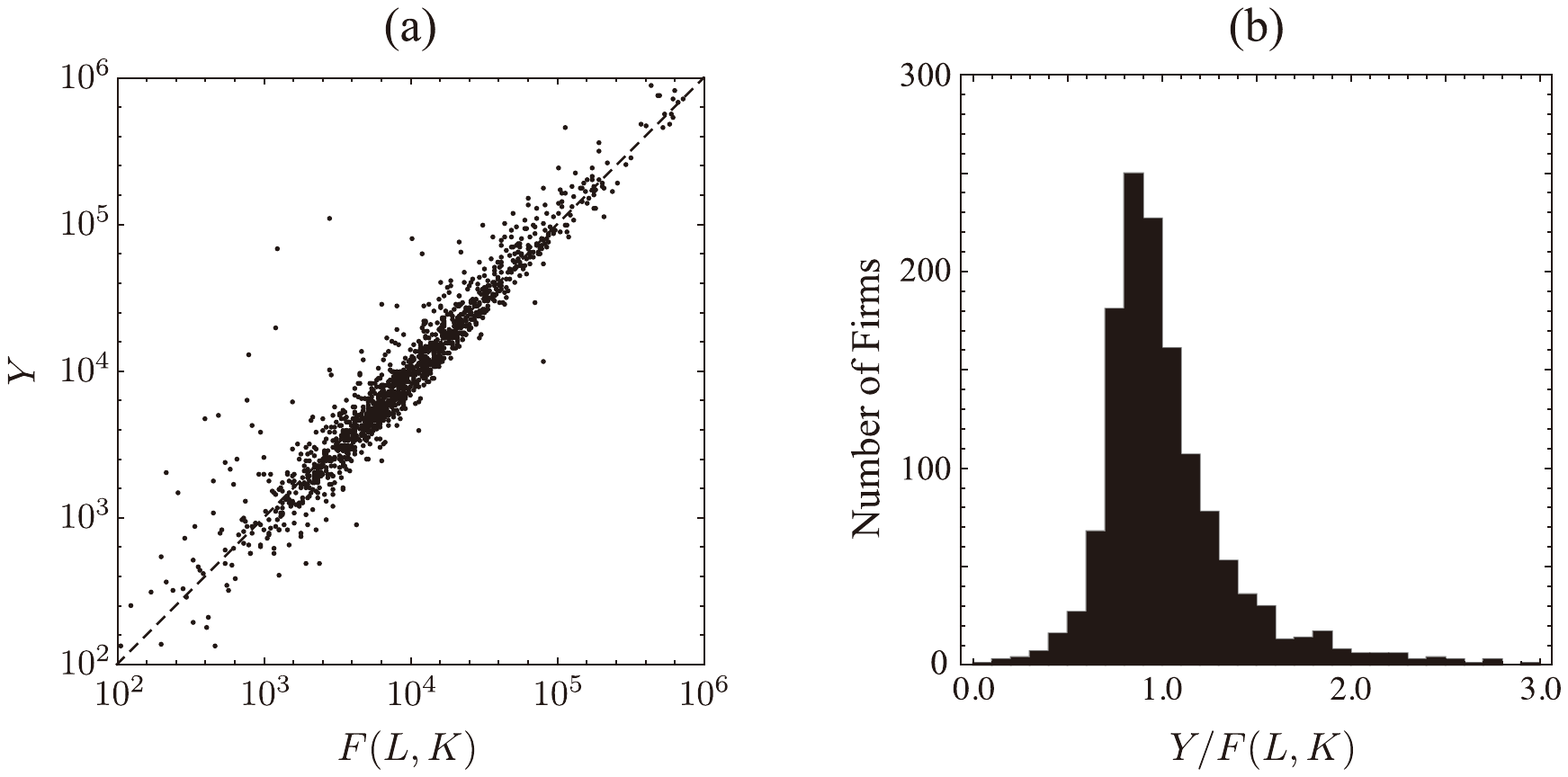}
\caption{The same as Fig.~\ref{fig:yycd}, but for the CES form.}
\label{fig:yyces}
\end{center}
\end{figure}

Adoption of the CES form, more flexible than the CD form, does not cure this failure of the production function at all. The CES form is given by
\begin{equation}
F(L, K)=A(\gamma L^{cp}+(1-\gamma)K^{cp})^{1/p} \,.
\label{eq:PF_CES}
\end{equation}
In the limit of $p \rightarrow 0$ this expression reduces to the CD form (\ref{eq:PF_CD}). We likewise determined the optimized set of the CES parameters by fitting Eq.~(\ref{eq:PF_CES}) to the real data:
\begin{equation}
A=1.617,\quad \gamma=0.932, \quad p=1.172, \quad c=1.004\; .
\label{eq:params_CES}
\end{equation}
The optimized CES function shares the {\it Constant Returns to Scale} property with the CD function, because the parameter $c$ so obtained is nearly equal to 1. Then we iterated the same calculations as previously carried out for the CD function. The results are depicted in Figs.~\ref{fig:pf_CES} and \ref{fig:yyces} corresponding to Figs.~\ref{fig:pf_CD} and \ref{fig:yycd}, respectively. We find no critical change in the conclusion which has been just drawn on the limited capability of the production function. In the CES form, quantitatively, 40\%(24\%) of 576 firms are outside the less-than-30\%(50\%)-discrepancy range in the case of $Y/F(K,L) >1$ and 16\%(4.0\%) of 784 firms are outliers for the same discrepancy range when $Y/F(K,L) <1$.

We thus need to work out an alternative theoretical device that allows us to take explicit account of the distribution itself, which we will discuss in
the following sections.

\section{Copula Theory}

A natural way to incorporate the stochastic nature of $Y$ even at given $K$ and $L$ is to construct a PDF model for the three financial quantities. Namely, the statistical distributions of the explanatory variables $L$, $K$ and the explained variable $Y$ in the production function are treated on an equal footing using the PDF.

The PDF is determined empirically according to
\begin{equation}
p\left( {L,K,Y} \right) = \frac{1}{N}\sum\limits_{i = 1}^N {\delta \left( {L - L_i } \right)\delta \left( {K - K_i } \right)\delta \left( {Y - Y_i } \right)},
\label{eq:emp_PDF}
\end{equation}
where the quantities with a subscript $i$ are those of a firm $i$ and $\delta (x) $ denotes Dirac's delta function. The empirical PDF (\ref{eq:emp_PDF}) is used as an estimator for a theoretical PDF. The CDF corresponding to $p\left( {L,K,Y} \right)$ is then given by
\begin{equation}
P_{<}\left( {L,K,Y} \right) = \int_0^L {\int_0^K {\int_0^Y {dL'dK'dY'}\, p\left( {L',K',Y'} \right)} } .
\label{eq:def_cd}
\end{equation}
The marginal PDF's with two variables are deduced from $p(L,K,Y)$ by integrating over one of the variables, e.g.,  
\begin{equation}
p\left( {L,Y} \right) = \int_0^\infty  {dK\,p\left( {L,K,Y} \right)}\ .
\label{eq:p3_to_p2}
\end{equation}
This relation is cast onto the CDF's such as
\begin{equation}
P_{<}\left( {L,Y} \right) = \int_0^L \int_0^Y {dL' dY'}\, p\left( {L',Y'} \right) = P_{<}\left( {L,K=\infty,Y} \right)\ .
\label{eq:cd3_to_cd2}
\end{equation}
Then the marginal PDF for each variable is likewise deduced from the binary PDF's as  
\begin{equation}
p\left( Y \right) = \int_0^\infty  {dL\,p\left( {L,Y} \right)} \, ,
\label{eq:PDF1_PDF2}
\end{equation}
for instance. Corresponding to Eq. (\ref{eq:PDF1_PDF2}), the marginal CDF is related to the binary CDF through  
\begin{equation}
P_{<}(Y)=\int_0^Y {dY'} p({Y'}) = P_{<}\left( {L=\infty,Y} \right) .
\end{equation}

In the present framework, unlike the production function (\ref{yf}), $Y$ is not determined uniquely
as a function of $K$ and $L$, but obeys a certain PDF. The PDF of $Y$ at a given set of $L$ and $K$ is calculated as a conditional probability:
\begin{equation}
p(Y|L,K) = \frac{{p\left( {L,K,Y} \right)}}{p\left( {L,K} \right)}\ .
\label{eq:PDF_Y}
\end{equation}
This describes the data exactly in a probabilistic way. Our task is thereby to adopt
a mathematical measure suitable for describing $p\left( {L,K,Y} \right)$, which we will do next.

\subsection{Definition of copula}

To work out an analytic model for $p\left( {L,K,Y} \right)$, we take advantage of a copula method. Copula is statistical means to measure dependence among stochastic variables. In other words, copulas extract correlations inherent among stochastic variables free from the marginal CDF's of variables themselves. A number of workable forms for copulas have been proposed by statisticians along with elucidation of their mathematical properties. 

Sklar's theorem~(\cite{sklar1959, Nelsen2006}) guarantees that the CDF $P_{<}( {L,K,Y} )$ is a unique function of the marginal CDF's associated with $L$, $K$, and $Y$:
\begin{equation}
P_{<}( {L,K,Y} ) = C( {u_L,u_K,u_Y} )\, ,
\end{equation}
where
\begin{equation}
u_{s} := P_{<}(s)=1-P_{>}(s)\quad (s = L, K, Y)\, ,
\end{equation}
are assumed to be continuous functions of $s$.
The function $C( {u_L,u_K,u_Y} )$ is called copula.
The PDF, $p( {L,K,Y} )$ is then derived from $P_{<}(L,K,Y)$ by carrying out partial differentiation with respect to each of the variables:
\begin{equation}
p( {L,K,Y} ) = \frac{{\partial ^3 P_{<}( {L,K,Y} )}}{{\partial L\partial K\partial Y}} = p(L) p(K) p(Y) c\left( {u_L ,u_K ,u_Y } \right),
\label{eq:PDF3}
\end{equation}
where we introduced the copula density $c( {u_L ,u_K ,u_Y })$ defined by
\begin{equation}
c\left( {u_L ,u_K ,u_Y } \right) := \frac{{\partial ^3 C\left( {u_L ,u_K ,u_Y } \right)}}{{\partial u_L \partial u_K \partial u_Y }}.
\end{equation}
If the variables are statistically independent of each other, then the copula density reduces to 1. Thus the copula density, referred to as the correlation function in many-body physics~(\cite{HM2006}), and hence the copula describe genuine correlations among the variables.

The copulas have the boundary conditions exemplified as
\begin{equation}
C\left( {u_L,u_K=0,u_Y } \right) = 0\ ,
\label{eq:copula3_bc1}
\end{equation}
\begin{equation}
C\left( {u_L ,u_Y } \right) = C\left( {u_L,u_K=1,u_Y } \right)\ .
\label{eq:copula3_bc2}
\end{equation}
The condition (\ref{eq:copula3_bc1}) is transparent from the definition (\ref{eq:def_cd}) for the CDF; Eq.~(\ref{eq:copula3_bc2}) just rephrases the relation (\ref{eq:cd3_to_cd2}) (\cite{Nelsen2006}). Also we note that if the financial quantities are independent of each other, then the copulas read,
\begin{equation}
C( {u_{s},u_{s'}} ) = u_{s}u_{s'},
\label{eq:cp2id}
\end{equation}
\begin{equation}
C( {u_{L},u_{K},u_{Y}} ) = u_{L}u_{K}u_{Y}.
\label{eq:cp3id}
\end{equation}

\subsection{Archimedean copulas}
There is a well-known family of copulas called Archimedean copulas (\cite{Nelsen2006}), which have been widely used because of their ease of mathematical handling together with diversity of their correlation properties. We first discuss the Archimedean copulas with two variables and then proceed to those with many variables. 

A bivariate Archimedean copula can be readily constructed from a generator function $\eta \left( {u} \right)$ as
\begin{equation}
C_{\rm{A}}\left( {u_1 ,u_2 } \right) = \eta ^{ - 1} \left( {\eta \left( {u_1 } \right) + \eta \left( {u_2 } \right)} \right) \, .
\label{eq:Arch_cop2}
\end{equation}
Here we assume that $\eta (u)$ is a continuous and strictly decreasing function mapping [0,1] to [0,$\infty$] with $\eta (1)=0$ so that its inverse $\eta ^{ - 1} (t)$ is well-defined for $0 \leq t \leq \infty$. We also have to impose convexity on $\eta (u)$ so that the constructed function (\ref{eq:Arch_cop2}) is a valid copula. 

In this paper we test three typical Archimedean copulas to model the real data. One of them is the Frank copula given by
\begin{align}
\eta_{\,\rm{F}} ( u;\theta) &= -\log \left( {\frac{{e^{ - \theta u}  - 1}}{{e^{ - \theta }  - 1}}} \right)\quad \mathrm{with}\, -\infty \leq \theta \leq \infty \, (\theta \neq 0) \, ,
\label{eq:cop2-Frank-genfnc}
\\
C_{\rm{F}} \left( {u_1,u_2; \theta} \right) &=  - \frac{1}{\theta }\log \Biggl[ {1 + \frac{{\left( {e^{ - \theta u_1 }  - 1} \right)} {\left( {e^{ - \theta u_2 }  - 1} \right)}}{{\left( {e^{ - \theta }  - 1} \right)}}} \Biggr].
\label{eq:cop2-Frank}
\end{align}
The second one is referred to as the Gumbel copula:
\begin{align}
\eta_{\,\rm{G}} ( u; \theta) &= \left( { - \log u} \right)^{\theta }\quad \mathrm{with}\; 1 \leq \theta \leq \infty,
\label{eq:cop2-Gumbel-genfnc}
\\
C_{\rm{G}} \left( {u_1,u_2; \theta } \right) &= \exp \left[ { - \left\{ {\left( { - \log u_1 } \right)^\theta   + \left( { - \log u_2 } \right)^\theta  } \right\}^{{1 \mathord{\left/
 {\vphantom {1 \theta }} \right.
 \kern-\nulldelimiterspace} \theta }} } \right].
\label{eq:cop2-Gumbel}
\end{align}
The last one is the Clayton copula given\footnote{For $\theta < 0$, $\eta_{\,\rm{C}}^{ - 1} (t)$ should be replaced by the pseudo-inverse $\eta_{\,\rm{C}}^{[- 1]} (t)$ which is equal to $\eta_{\,\rm{C}}^{ - 1} (t)$ in $0 \leq t \leq \eta_{\,\rm{C}}(0)$ but set to be 0 beyond $t = \eta_{\,\rm{C}}(0)$.} by
\begin{align}
\eta_{\,\rm{C}} ( u; \theta) &= \frac{1}{\theta} \left( u^{ - \theta }  - 1 \right)\quad \mathrm{with}\, -1 \leq \theta \leq \infty \, (\theta \neq 0) \, ,
\label{eq:cop2-Clayton-genfnc}
\\
C_{\rm{C}} \left( {u_1 ,u_2; \theta} \right) &=\left[ \max \left( {u_1 ^{ - \theta }  + u_2 ^{ - \theta }  - 1},0 \right) \right]^{-{1 / \theta }}.
\label{eq:cop2-Clayton}
\end{align}

\begin{figure}[!tp]
\begin{center}
\includegraphics[width=0.9\textwidth,bb=0 0 1697 563]{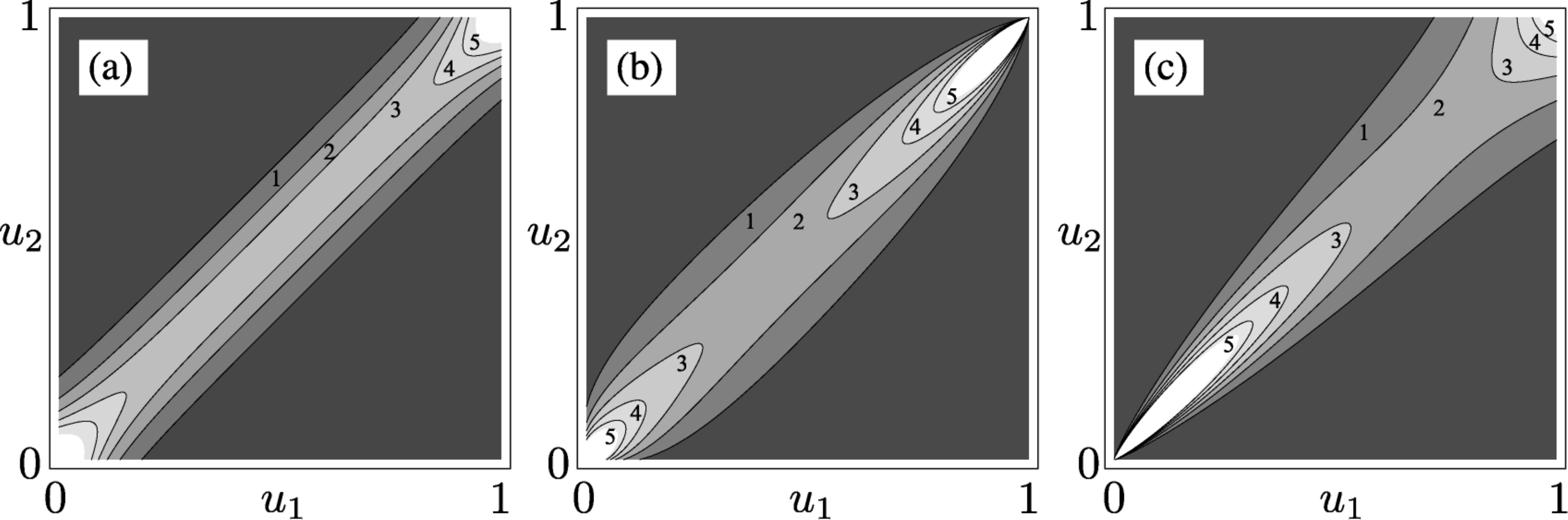}
\caption{Contour plots of the bivariate copula densities. The panel (a) is for Frank with $\theta=14.14$; (b), for Gumbel with $\theta=4$; (c), for Clayton with $\theta=6$. The parameter $\theta$ is chosen to yield the same value of the Kendall $\tau\,(= 0.75)$ for each copula. Level values of the contours are annotated on the figures.}
\label{copula2_density_FGC}
\end{center}
\end{figure}

These three copulas possess correlation properties of different characteristics as displayed in Fig.~\ref{copula2_density_FGC}. The strength of dependence in the Frank copula is almost flat spanning from the lower tail ($u_1, u_2 \simeq 0$) to the upper tail ($u_1, u_2 \simeq 1$). The Gumbel copula has stronger correlation in the upper tail than in the lower tail and the Clayton copula shows reversed correlation structure.

This asymmetric nature in correlation structure of copulas is, however,
superficial. In fact, by changing the variables, $u_i \rightarrow 1-u_i$, we derive a different copula model from a given copula, called survival copula. For instance, the survival Clayton copula (s-Clayton, for short), denoted as $\hat{C}_{\rm{C}}\left( u_1,u_2; \theta \right)$, is related (\cite{Nelsen2006}) to the original Clayton copula through 
\begin{equation}
\hat{C}_{\rm{C}}\left( u_1,u_2; \theta \right)=u_1 + u_2 -1 + C_{\rm{C}}\left( 1-u_{1},1-u_{2}; \theta \right).
\label{eq:cop2-Clayton-survival}
\end{equation}
As is easily appreciated, in turn, $\hat{C}_{\rm{C}}( u_1,u_2)$ has stronger correlation in the upper tail.

The bivariate Archimedean copulas can be generalized to multivariate copulas with $n$ variables in such an iterative way as 
\begin{equation}
C_{\rm{A}}\left( {u_1 ,\cdots, u_n} \right) = \eta ^{ - 1} \left( {\eta \left( {u_1 } \right) + \cdots + \eta \left( {u_n } \right)} \right).
\label{eq:Arch_cop3}
\end{equation}
The generator function however is required to satisfy a more stringent mathematical constraint that its inverse $\eta^{-1}(x)$ is completely monotonic, i.e., 
\begin{equation}
(-1)^m \frac{d^m \eta^{-1}(x)}{dx^{m}} \geq 0 \quad(m=0,1,2,\cdots).
\label{eq:monotone}
\end{equation}
We refer the readers to the textbook by \cite{Nelsen2006} for the details. The generators of Frank, Gumbel, and Clayton are completely monotonic with the following conditions for $\theta$: $\theta > 0$ (Frank), $\theta > 1$ (Gumbel), and $\theta > 0$ (Clayton). We will call the multivariate Archimedean copula (\ref{eq:Arch_cop3}) in a more specific way after its generator; namely, the copula (\ref{eq:Arch_cop3}) generated by Eq.~(\ref{eq:cop2-Frank-genfnc}) is referred to as $n$-variate Frank copula and so forth.

\subsection{Generalization of Archimedean copulas}
The Archimedean copulas, widely used for applications, may be sometimes too restrictive to accommodate real data.

They are totally symmetric with respect to exchange of independent variables. A way to relax this constraint for the bivariate copulas is to set 
\begin{equation}
C_{{\rm{A}}} ( {u_1 ,u_2 }; \theta, \alpha, \beta ) = u_1^{1 - \alpha } u_2^{1 - \beta } C_{\rm{A}} ( {u_1^\alpha  ,u_2^\beta  }; \theta ),
\label{eq:cop2-Gumbel-asym}
\end{equation}
where two additional parameters with $0 \leq \alpha, \beta \leq 1$ are introduced. In the case of $\alpha = \beta$ the generalized copula is exchangeable. Difference between $\alpha$ and $\beta$ thus measures the degree of asymmetry in correlations. Taking a special limit of $\alpha = \beta = 1$ recovers the original Archimedean copulas.

Also we recall that the trivariate Archimedean copula, Eq. (\ref{eq:Arch_cop3}) with $n=3$, is characterized by a single parameter $\theta$. It means that all of the marginal CDF's have the same correlation structure. This is not true for the real data under study as has been already indicated, so that we generalize the original form as
\begin{equation}
\begin{split}
C_{\textrm{A-nex}}( {u_1 ,u_2, u_3}; \theta_{1}, \theta_{2}) 
&= C_{\textrm{A}}( C_{\textrm{A}}(u_{1}, u_{2};\theta_{2}), u_{3}; \theta_{1}) \\
&= \eta_{1} ^{ - 1}
\left( {\eta_{1} \left( {\eta_{2} ^{ - 1} \left( {\eta_{2} \left( {u_1 } \right) + \eta_{2} \left( {u_2 } \right)} \right)} \right) + \eta_{1} \left( {u_3 } \right)} \right).
\end{split}
\label{eq:Arch-asym_cop3}
\end{equation}
with different generator functions $\eta_{1}(u) = \eta(u; \theta_{1}) $ and $\eta_{2}(u) = \eta(u; \theta_{2})$. The copula (\ref{eq:Arch-asym_cop3}) now has two characteristic parameters $\theta_1$ and $\theta_2$ leading to the margin $C_{\textrm{A}}( {u_1 ,u_2}; \theta_{2})$ generated by $\eta_{2}(u)$ and the remaining margins, $C_{\textrm{A}}( {u_1 ,u_3}; \theta_{1})$ and $C_{\textrm{A}}( {u_2 ,u_3}; \theta_{1})$ generated by $\eta_{1}(u)$. If one set $\theta_{1}=\theta_{2}$ in Eq. (\ref{eq:Arch-asym_cop3}), the exchangeable Archimedean copula (\ref{eq:Arch_cop3}) with $n=3$ is recovered. We note that the condition $\theta_{1} \leq \theta_{2}$ should be satisfied for Eq.~(\ref{eq:Arch-asym_cop3}) to be a copula. We refer to the generalized form~(\ref{eq:Arch-asym_cop3}) as non-exchangeable Archimedean copula following \cite{MFE2005}.

\section{Construction of Production Copula}
In this section the distributed real data are modeled using copulas as they are. We refer to the resulted copula as ``production copula".

\subsection{Bivariate}
First we focus on the bivariate correlations for all pairs out of $L$, $K$, and $Y$ as depicted in Fig.~\ref{KLY2_NEEDS2006}. Three Archimedean copulas due to Frank, Gumbel and Clayton are examined as has been mentioned. Actually the survival copula of Clayton is used instead of the original one, because the former is better suited to description of the correlation structure in the real data than the latter.

\begin{table}
\caption{Maximized log-likelihood $\ell$ in fitting of bivariate Archimedean copulas (Frank, Gumbel, s-Clayton) to the pair correlations in the real data as shown in Fig.~\ref{KLY2_NEEDS2006}. Also the Kendall rank correlation coefficient $\tau$ corresponding to the value of $\theta$ is listed for each copula.}
\begin{center}
\begin{tabular}{|ll||c|c|c|}
\hline
Copula & & $K$-$Y$ & $L$-$Y$ & $L$-$K$ \\
\hline
\hline
Frank & $\theta$ & 11.0 & 21.2 & 9.23\\
      & $\ell$ & 940.1 & 1612.4 & 790.2 \\
      & $\tau$ & 0.691 & 0.826 & 0.644 \\
\hline
Gumbel & $\theta$ & 3.21 & 5.30 & 2.73 \\
       & $\ell$ & 1081.8 & 1694.0 & 892.1\\
       & $\tau$ & 0.688 & 0.811 & 0.634 \\
\hline 
s-Clayton & $\theta$ & 3.43 & 6.11 & 2.59 \\
                   & $\ell$ & 992.1 & 1483.7 & 787.3 \\
                   & $\tau$ & 0.632 & 0.753  &  0.564\\
\hline
\end{tabular}
\end{center}
\label{tab:mlm_fit_spf}
\end{table}

\begin{figure}[!tp]
\begin{center}
\includegraphics[width=\textwidth,bb=51 340 544 501]{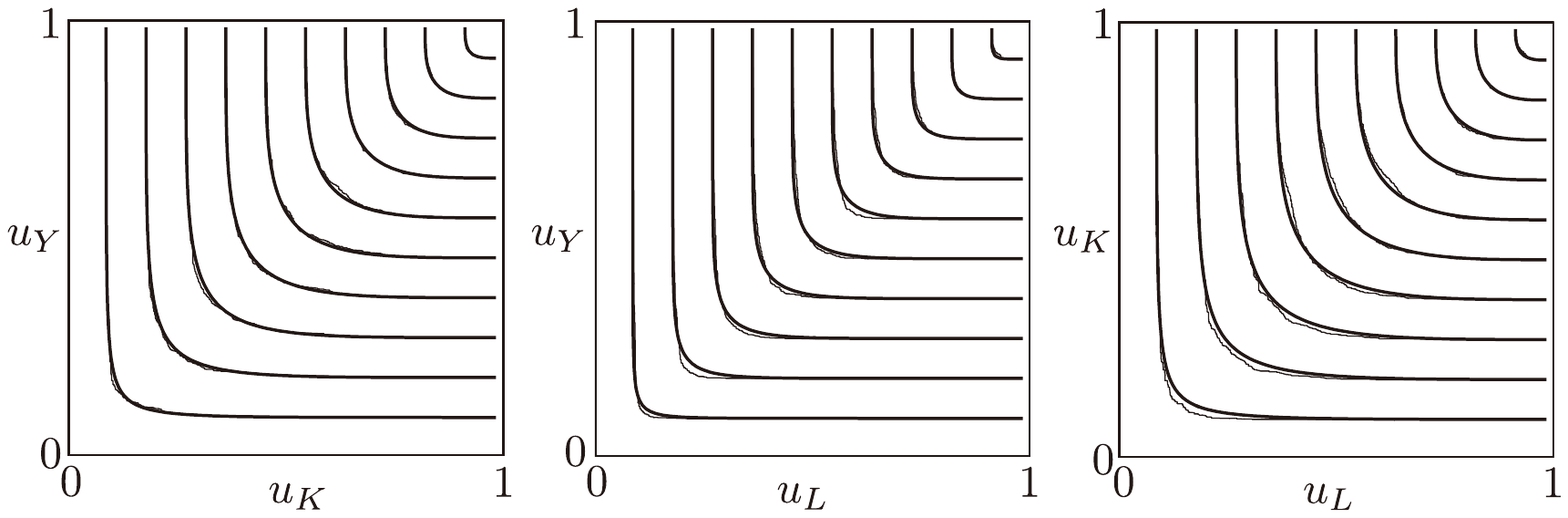}
\caption{Modeling of the pair correlations in Fig.~\ref{KLY2_NEEDS2006} in terms of the Gumbel copula. The fitted results (smooth solid curves) are compared with the corresponding real data (jaggy solid curves); difference between the two curves is almost invisible. Ten contours are drawn at equal spacing ranging from 0 (on the bottom horizontal axis and the left vertical axis) to 1 (at the top right corner) on each panel.}
\label{copula2_Gumbel}
\end{center}
\end{figure}

\begin{figure}[!tp]
\begin{center}
\includegraphics[width=\textwidth,bb=109 0 2055 562]{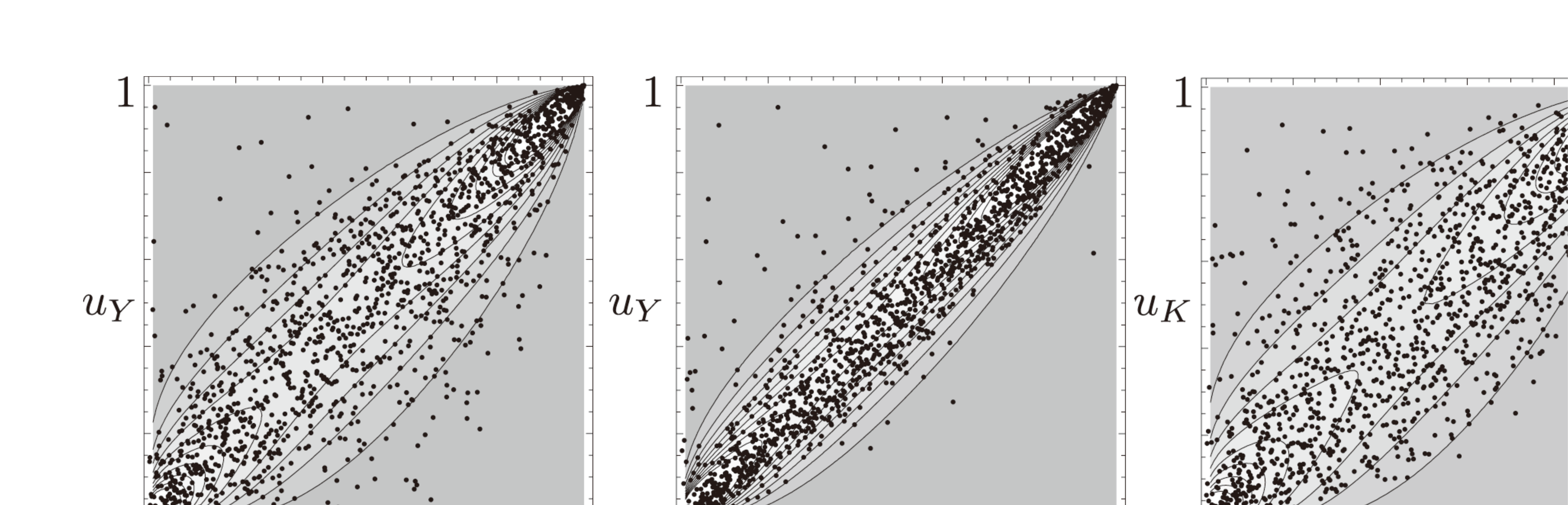}
\caption{Contour plots of the copula densities corresponding to the fitted copulas in Fig.~\ref{copula2_Gumbel}. The dots refer to the real data. The contours are equally spaced levels with the interval of 0.5 ranging from 0.5 (dark side) to 5 (bright side).}
\label{copula2_density_Gumbel}
\end{center}
\end{figure}

We have carried out the maximum likelihood estimate to fit those copulas to the pair correlations as given in Fig.~\ref{KLY2_NEEDS2006}. The fitting results are summarized in Table~\ref{tab:mlm_fit_spf}, where the optimized values of the correlation parameter $\theta$ and the maximized log-likelihood $\ell$ are listed. Also we calculated the Kendall rank correlation coefficient $\tau$ from the $\theta$'s for each copula. We observe that the Gumbel copula gives the best fitting among the three copulas tested as manifested by the largest value for $\ell$. The optimized results with the Gumbel copula are thus compared with the corresponding empirical copulas derived from the real data in Fig.~\ref{copula2_Gumbel}. The two results are in good agreement with each other. The same comparison but for the copula densities are also made in Fig.~\ref{copula2_density_Gumbel}.

The overall strength of the pair correlations is represented by magnitude of $\theta$ and $\tau$. Relative comparison of $\theta$'s and $\tau$'s in Table~\ref{tab:mlm_fit_spf} confirms the conclusion drawn in the previous section. The $L$-$Y$ correlation is significantly stronger than the $K$-$Y$ and $L$-$K$ correlations. In contrast, the latter two correlations resemble each other. Furthermore, we address to what extent the bivariate correlations among $L$, $K$, and $Y$ are asymmetric. We repeated the maximum likelihood estimate using the asymmetric Gumbel copula given by Eq.~(\ref{eq:cop2-Gumbel-asym}) with Eq.~(\ref{eq:cop2-Gumbel-genfnc}). Table~\ref{tab:mlm_fit_tpf} demonstrates that none of the pairs has notable asymmetry in its correlation structure.

\begin{table}
\caption{Degree of asymmetry in the pair correlations in Fig.~\ref{KLY2_NEEDS2006} estimated using the asymmetric Gumbel copula.}
\begin{center}
\begin{tabular}{|c||c|c|c|}
\hline
 & $K$-$Y$ & $L$-$Y$ & $L$-$K$ \\
\hline
\hline
$\theta$& 3.34 & 6.59 & 2.86 \\
\hline
$\alpha$ & 0.994 & 0.999 & 1.000 \\
\hline 
$\beta$& 0.984 & 0.934 & 0.957 \\
\hline
$\ell$& 1090.1 & 1779.5 & 896.0 \\
\hline
\end{tabular}
\end{center}
\label{tab:mlm_fit_tpf}
\end{table}

In passing, we recall copulas based on Gaussian and Student's $t$ distributions~(\cite{Cherubini2004}, \cite{MS2006}), which are also popular in practical applications as well as Archimedean copulas. However, those copulas have no simple closed forms such as Archimedean copulas have. The bivariate Gaussian copula is characterized by a single parameter $\zeta$ that is the correlation coefficient. We reiterated modeling of the real data using the Gaussian copula, and the results of maximum likelihood estimate are summarized in Table~\ref{tab:fit_gt}. Comparison with the corresponding results in Table~\ref{tab:mlm_fit_tpf} shows that the Gumbel copula is superior to the Gaussian copula over all pairs of the financial quantities. The bivariate Student's $t$ copula has one more parameter $m$, the number of degrees of freedom, in addition to the correlation coefficient $\zeta$; the limit of $m \rightarrow \infty$ reduces the $t$ copula to the Gaussian copula. The maximized log-likelihood $\ell$ obtained with the $t$ copula is also listed for each of the pairs in Table~\ref{tab:fit_gt}, where a typical value of $m$, i.e., $m=3$ was adopted. We see that the Gumbel copula still works better than the $t$ copula except for the $L$-$Y$ pair. This is a quantitative reason why we prefer Archimedean copulas in this study, although the Gumbel copula might be superseded by the $t$ copula with more appropriate choice of $m$. We also respect the simplicity of Archimedean copulas.

\begin{table}
\caption{Maximum likelihood estimate for bivariate Gaussian and Student's $t$ ($m=3$) copulas fitted to the real data in Fig.~\ref{KLY2_NEEDS2006}.}
\begin{center}
\begin{tabular}{|cl||c|c|c|}
\hline
Copula & & $K$-$Y$ & $L$-$Y$ & $L$-$K$ \\
\hline
\hline
Gaussian & $\zeta$ & 0.868 & 0.945 & 0.844\\
      & $\ell$ & 970.0 & 1538.3 & 864.0 \\
\hline
$t$    & $\zeta$ & 0.874 & 0.959 & 0.830\\
       & $\ell$ & 1050.9 & 1749.5 & 887.4 \\
\hline 
\end{tabular}
\end{center}
\label{tab:fit_gt}
\end{table}

\subsection{Trivariate}

At last we are ready to discuss the principal objective of the present paper, namely, construction of the production copula. The models we consider here are enumerated as
\begin{description}
\item[Model (I)] Frank copula; 
\begin{equation}
C^{\rm (I)}(u_L, u_K, u_Y)=C_{\rm F} (u_L, u_K, u_Y; \theta).
\end{equation}
\item[Model (II)] Gumbel copula; 
\begin{equation}
C^{\rm (II)}(u_L, u_K, u_Y)=
C_{\rm G} (u_L, u_K, u_Y; \theta).
\end{equation}
\item[Model (III)] s-Clayton copula;
\begin{align}
C^{\rm (III)}(u_L, u_K, u_Y)&=\hat{C}_{C}(u_L, u_K, u_Y ; \theta)\nonumber\\
&=u_L + u_K + u_Y -1 + C_{\rm C} (1-u_L, 1-u_K, 1-u_Y; \theta).
\end{align}
\item[Model (IV)] Non-exchangeable Gumbel copula;
\begin{equation}
C^{\rm (IV)}(u_L, u_K, u_Y)=
C_{\rm G} (C_{\rm G} (u_L, u_Y; \theta_2), u_K; \theta_1).
\end{equation}
\end{description}
In the above, trivariate Archimedean copulas $C_{\rm F,G,C} (u_L, u_K, u_Y; \theta)$
are constructed from their $\eta$-functions 
(\ref{eq:cop2-Frank-genfnc}), (\ref{eq:cop2-Gumbel-genfnc}), (\ref{eq:cop2-Clayton-genfnc})
respectively by the prescription (\ref{eq:Arch_cop3}).
The first three models specified by a single parameter result in all of marginal CDF's and PDF's with the same correlation structure. The last model is based on Eq.~(\ref{eq:Arch-asym_cop3}) having Eq.~(\ref{eq:cop2-Gumbel}) plugged into $C_{\textrm{A}}(u_{i}, u_{j};\theta_{1})$ and $C_{\textrm{A}}(u_{i}, u_{j};\theta_{2})$. The variables $u_1$, $u_2$, and $u_3$ in Eq.~(\ref{eq:Arch-asym_cop3}) should read $u_L$, $u_Y$, and $u_K$, respectively. The last model can account for the difference in correlations among the marginals with different values for $\theta_{1}$ and $\theta_{2}$, which is a desirable property as we will see. 

\begin{table}[htdp]
\caption{Maximum likelihood optimization in the models for the production copula.}
\begin{center}
\begin{tabular}{|c||c|c|c|c|}
\hline
 & Model (I) & Model (II) & Model (III) & Model (IV) \\
\hline
\hline
$\theta$ & 11.5  & 3.27 & 3.34 & 2.89 ($\theta_1$), 5.26 ($\theta_2$) \\
\hline
$\ell$ & 2207.5 & 2428.2  & 2164.3 & 2701.1\\
\hline
$AIC$ & -4413.0 & -4854.4  & -4326.6 & -5398.2 \\
\hline 
\end{tabular}
\end{center}
\label{tab:mlm_fit_cumulant3_KLY}
\end{table}

We adapted the four models to the real data using the MLM again. Results of the optimization are listed in Table~\ref{tab:mlm_fit_cumulant3_KLY}. As has been easily expected from the experience in modeling the bivariate correlations, the Gumbel copula used in Model (II) certainly outperforms the fitting as compared with the other Archimedean copulas in Models (I) and (III). This is actually the motivation behind Model (IV) in which the Gumbel copula is specially selected. The result of the optimization based on the generalized model is also included in Table~\ref{tab:mlm_fit_cumulant3_KLY}. Since the numbers of parameters are different between Models (II) and (IV), we have to replace the maximum log-likelihood $\ell$ by Akaike's information criterion $AIC$ for the model selection (\cite{akaike1974}); a model with the smallest $AIC$ should be adopted. The criterion is given by
\begin{equation}
AIC = -2\ell+2k,
\label{eq:AIC}
\end{equation}
where $k$ is the number of parameters in a statistical model. Comparison of the $AIC$ values proved that Model (IV) significantly improves the fitting over even Model (II).

Further we delve into the performance of Model (IV). To make detailed comparison with the real data, we introduce a trivariate copula cumulant\footnote{The cumulant functions are usually defined in terms of the PDF's in place of the CDF's (\cite{HM2006}).} defined by
\begin{align}
\Omega(u_{L},u_{K},u_{Y}) 
&= C(u_{L},u_{K},u_{Y}) - u_{K}C(u_{L},u_{Y}) - u_{L} C(u_{K},u_{Y}) \notag \\
&\quad - u_{Y}C(u_{L},u_{K}) + 2 u_{L}u_{K}u_{Y} \, .
\label{eq:cumulant3_KLY}
\end{align}
Here the contributions essentially due to the bivariate correlations are subtracted from the trivariate copula. To appreciate this fact, we will take two special cases. First suppose that all of the variables are independent of each other. Replacement of the copulas on the right-hand side of Eq. (\ref{eq:cumulant3_KLY}) by the corresponding independent copulas (\ref{eq:cp2id}) and (\ref{eq:cp3id}) leads to $\Omega(u_{L},u_{K},u_{Y}) = 0$. Next suppose that only a pair of $L$ and $K$ are intercorrelated among the three variables, then the copula $C(u_{L},u_{K},u_{Y})$ is decomposed as
\begin{equation}
C(u_{L},u_{K},u_{Y})=u_{Y}C(u_{L},u_{K}).
\label{eq:copula3id}
\end{equation}
And the bivariate copulas involving $u_{Y}$ as a variable can be replaced by the corresponding independent copulas. Again we see that $\Omega(u_{L},u_{K},u_{Y})$ vanishes in this case. Such subtraction is also manifested by the boundary condition that the copula cumulant vanishes on the marginal boundaries, namely,
\begin{equation}
\Omega(1,u_{K},u_{Y}) = \Omega(u_{L},1,u_{Y}) = \Omega(u_{L},u_{K},1) = 0.
\label{eq:cumulant3_KLY_earth}
\end{equation}

\begin{figure}[!tp]
\begin{center}
\includegraphics[width=0.8\textwidth,bb=103 38 529 798]{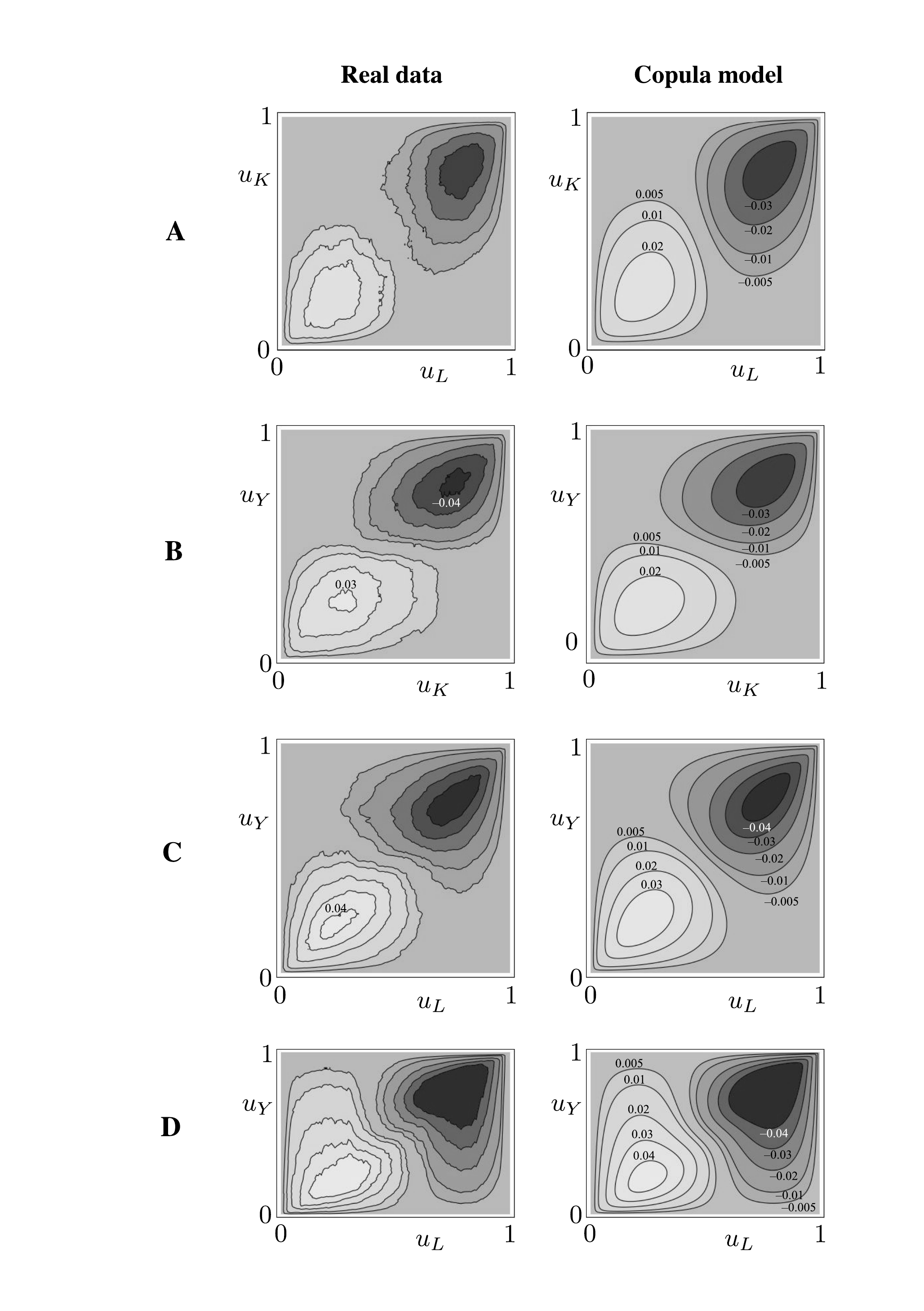}
\caption{Contour plots of the copula cumulant, Eq.~(\ref{eq:cumulant3_KLY}) on the cross sections in $u_L$-$u_K$-$u_Y$ space as specified in Fig.~\ref{fig:cross_sections}. The results obtained from the real data (left-hand side) are compared with those derived from the production copula in Model (IV) (right-hand side); the contours are drawn at the same levels on both sides.}
\label{fig:copula3_cross_sections}
\end{center}
\end{figure}

\begin{figure}[!tp]
\begin{center}
\includegraphics[width=0.6\textwidth,bb=95 278 492 557]{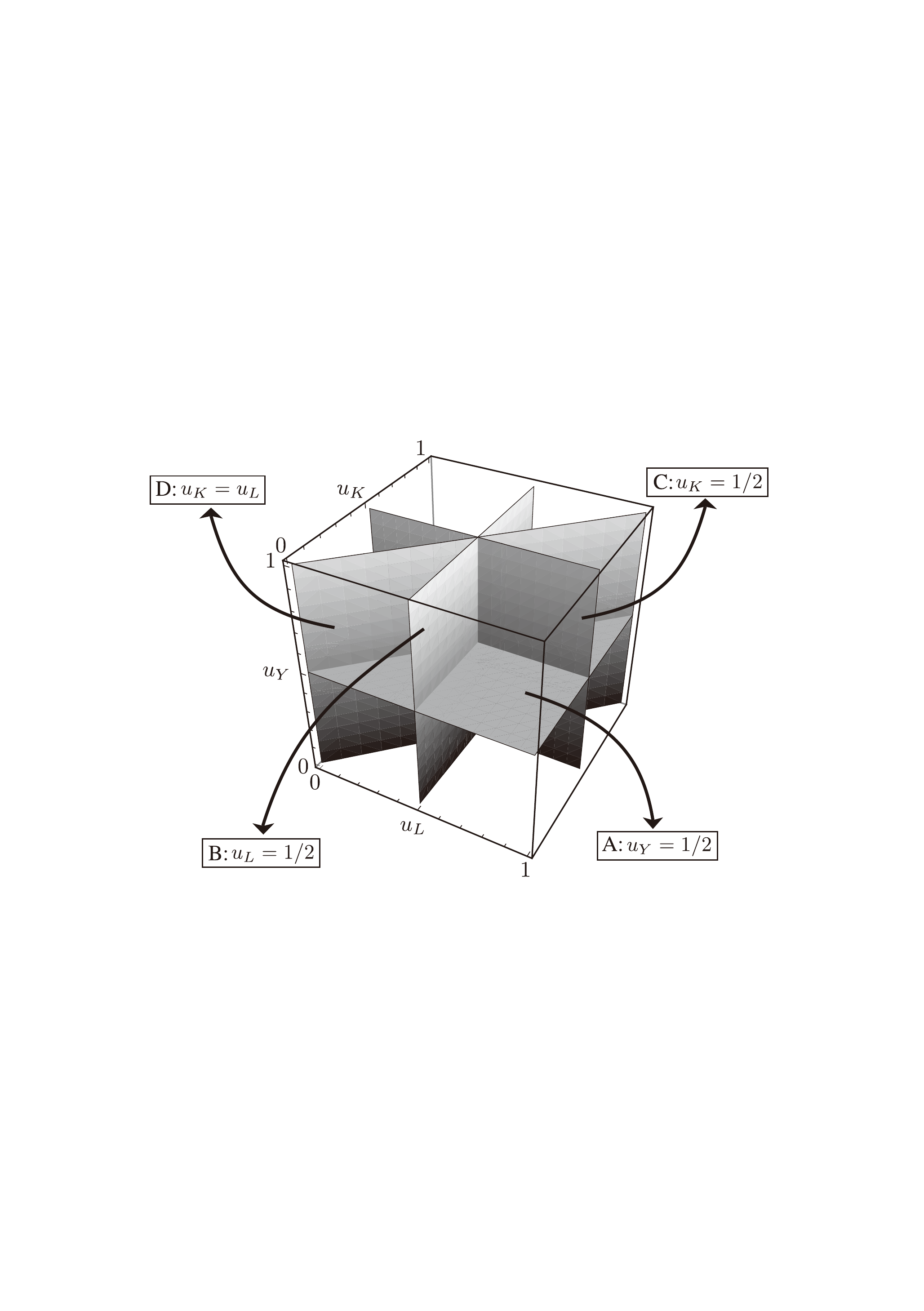}
\caption{Typical cross sections in $u_L$-$u_K$-$u_Y$ space.}
\label{fig:cross_sections}
\end{center}
\end{figure}

\begin{figure}[!tp]
\begin{center}
\includegraphics[width=0.5\textwidth,bb=163 344 415 664]{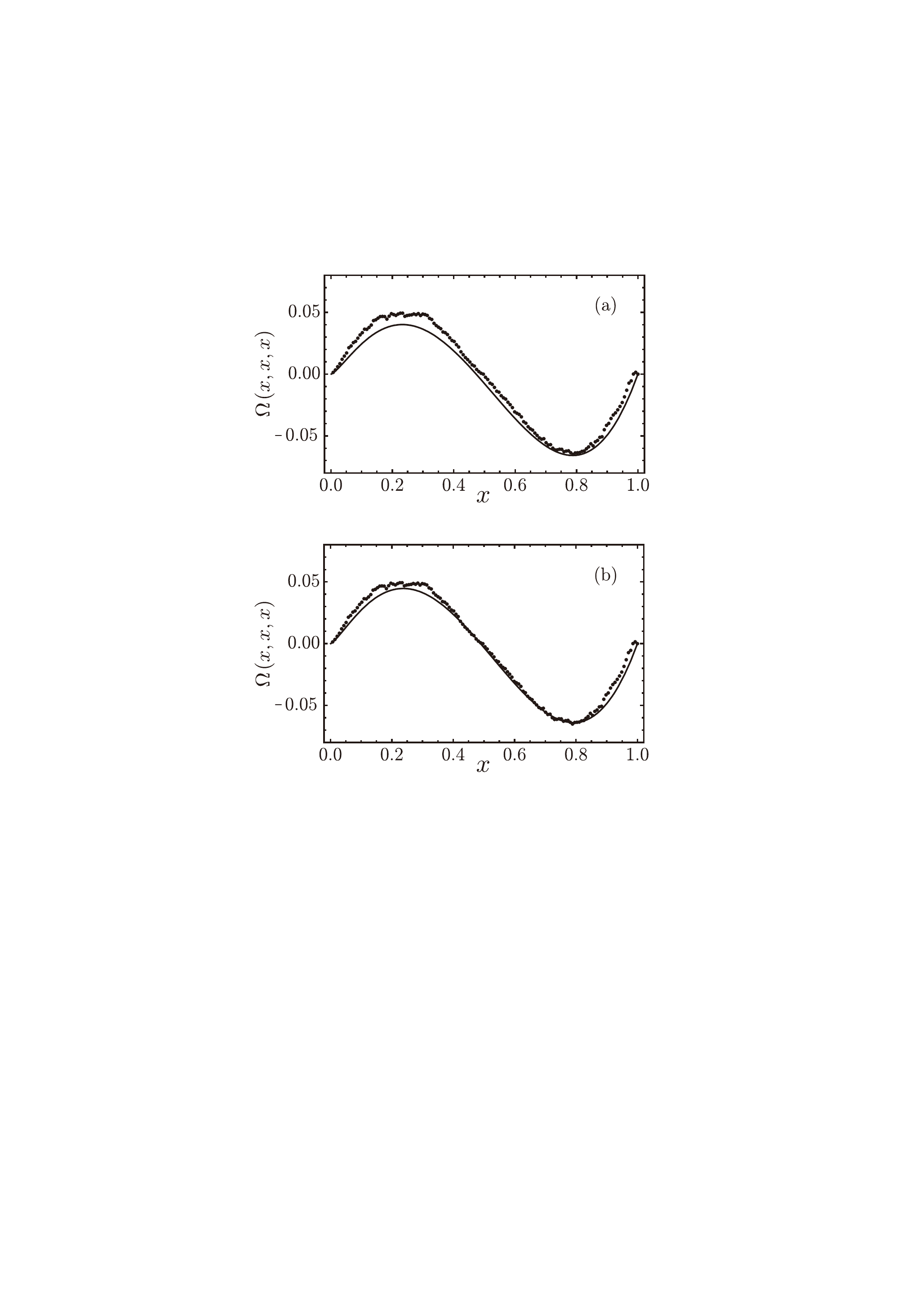}
\caption{Comparison of the results for $\Omega(x,x,x)$ between the copula model (solid curve) and the real data (dots). Model (II) is adopted in the panel (a); Model (IV), in the panel (b).}
\label{fig:copula3_Gumbel}
\end{center}
\end{figure}

Figure~\ref{fig:copula3_cross_sections} shows the fitted results for $\Omega(u_{L},u_{K},u_{Y})$ in Model (IV) together with the corresponding empirical data on the typical cross sections A-D depicted in Fig.~\ref{fig:cross_sections}. We observe the copula model reproduces the empirical results in an almost indistinguishable manner. Figure~\ref{fig:copula3_Gumbel} gives more in-depth comparison of the results between the copula model and the real data for $\Omega(x,x,x)$ along the diagonal direction specified by $x=u_{L}=u_{K}=u_{Y}$. Again switching from Model (II) to (IV) leads to significant improvement in reproducing the empirical copula cumulant.

\begin{figure}[!tp]
\begin{center}
\includegraphics[width=0.8\textwidth,bb=0 0 1952 2062]{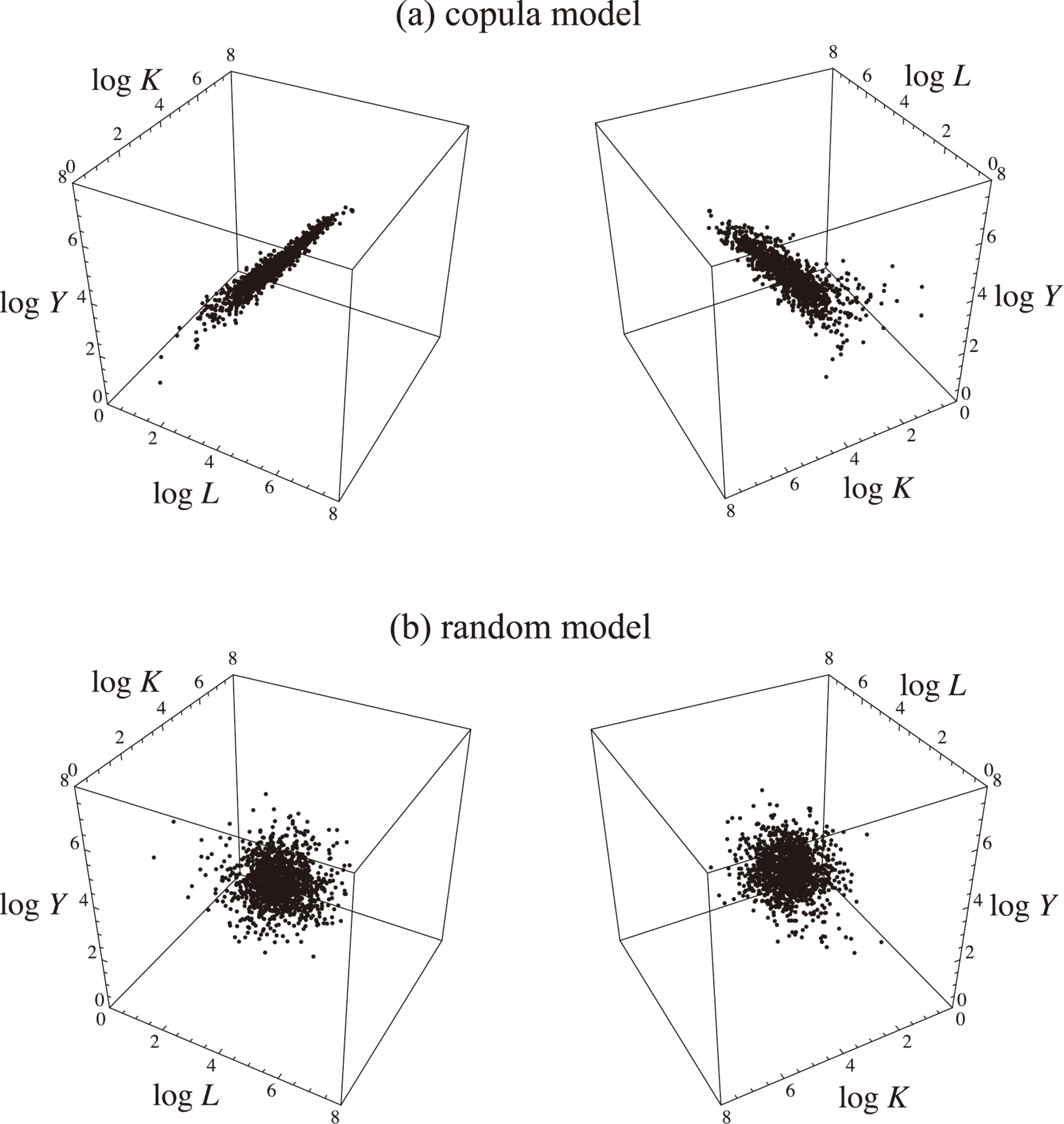}
\caption{A simulated result for the production activity of the Japanese listed firms obtained with the production copula in Model (IV), accompanied by that based on the random model without any correlations among the financial quantities. The 3D graphs are drawn with the same view angles as in Fig.~\ref{fig:pf_CD} for each model.}
\label{fig:copula3_model}
\end{center}
\end{figure}

Now that the copula model has been established, it can
be used for various economic studies of firms.
Here we demonstrate one such an example. We carried out a simulation for the production activity of the Japanese listed firms in 2006 and exhibit the result in Figure~\ref{fig:copula3_model}. The simulated points ($N$ = 1360) were generated using a rejection method~(\cite{NR2007}) with the trivariate PDF obtained by combining the production copula and the marginal PDF's according to Eq.~(\ref{eq:PDF3}). 
For reference we did a simulation based on the random model where no correlations among the financial quantities were taken into account, as also shown in Fig.~\ref{fig:copula3_model}. These results should be compared with the real data depicted in the same way as in Fig.~\ref{fig:pf_CD}. The comparison confirms how powerful is the production copula.

\begin{figure}[!tp]
\begin{center}
\includegraphics[width=0.9\textwidth,bb=55 291 527 513]{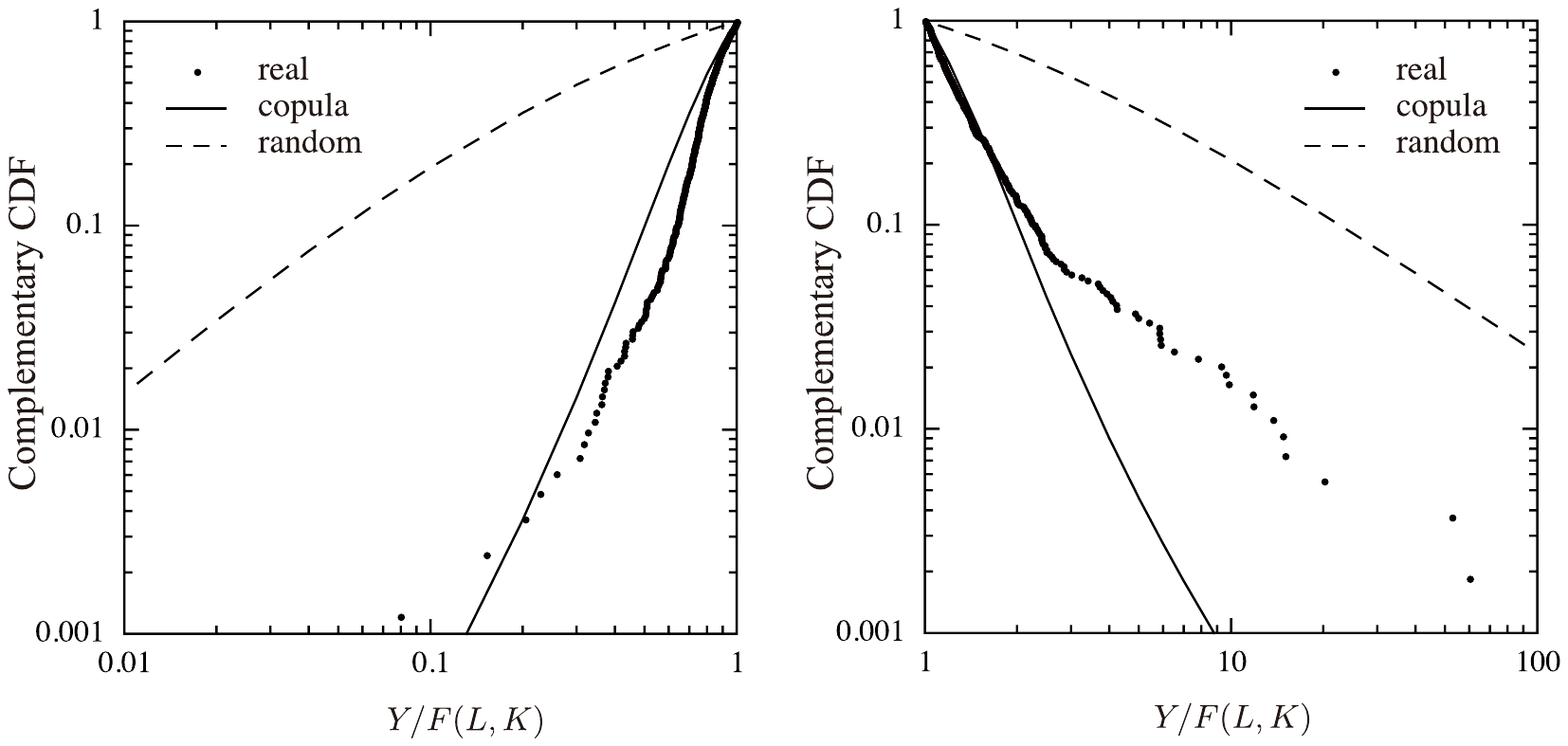}
\caption{Complementary CDF's of $Y/F(L,K)$ calculated in Model (IV), compared to the corresponding results based on the real data as shown in Fig.~\ref{fig:yycdfat} and those based on the random model as demonstrated in Fig. \ref{fig:copula3_model}.}
\label{fig:yycopulafat}
\end{center}
\end{figure}

Also we calculated the deviation from the CD production function for the copula and random models, as has been already done for the real data in Figs.~\ref{fig:yycd} and \ref{fig:yycdfat}. One can express the complementary CDF of the ratio $\xi=Y/F(L,K)$ on both sides in terms of the trivariate copula. For instance, the complementary CDF on the upper side is given by
\begin{equation}
P_{>}(\xi)=f(\xi)/f(1) \quad (1 \leq \xi < \infty)\, ,
\label{eq:ratio_Y_to_CD}
\end{equation}
where
\begin{equation}
f(\xi)=1-\int_0^1 du_{L} \int_0^1 du_{K} \frac{\partial^{2} C(u_{L}, u_{K}, u_{Y_{\xi}})}{\partial u_{L} \partial u_{K}}\, ,
\label{eq:ratio_Y_to_CD_num}
\end{equation}
with
\begin{equation}
Y_{\xi} = \xi F(L,K) = \xi F(P_{<}^{-1}(u_{L}),P_{<}^{-1}(u_{K}))\, .
\label{eq:Y_xi_CD}
\end{equation}
For the random model, Eq. (\ref{eq:ratio_Y_to_CD_num}) is replaced by
\begin{equation}
f(\xi)=1-\int_0^1 du_{L} \int_0^1 du_{K} u_{Y_{\xi}}\, .
\label{eq:ratio_Y_to_CD_num_random}
\end{equation}
Figure~\ref{fig:yycopulafat} makes detailed comparison of the results based on the copula and random models with the corresponding real data. The analytic model (\ref{bbdef}) for the marginal CDF's was highly useful in executing numerical computation of the double integration in Eq.~(\ref{eq:ratio_Y_to_CD_num}). We see that the correlations involved in the real data are well reproduced by the copula model, except for outliers on the upper side occupying about 5\% of the data; the functional behavior of them is rather close to that of the results in the random model.

To calculate $p\left( {L,K,Y} \right)$ in the present copula model, in fact, we are required to determine totally 14 parameters, 12 for the three marginal distributions and 2 for the copula. The reader may think that the production copula contains much more parameters than the CD production function which has just 3 parameters. This is not true at all, because we have to specify $p\left( {L,K} \right)$ to take account of correlations between the input variables for the production function. This is a fair way to compare the two ideas. If the beta distribution of the second kind and the Gumbel copula are likewise employed, totally 9 additional parameters appear, that is, 8 for the two marginal distributions and 1 for the copula. We thus claim that the production copula is so successful in reproducing the real data and so workable considering that it has only 2 extra parameters.

\section{Conclusion}
In this paper we have proposed the use of the production copula to take full account of a wide variety of production activities of firms.

We showed that capital, labor and value added of firms are closely related in a probabilistic way to each other by analyzing financial data of listed firms in Japan. At the same time we confirmed that the productive heterogeneity of firms was far beyond the scope of a production function. These empirical facts authorizes our endeavor to model the production activities of firms in terms of copulas. Four copula models were employed and their accuracy was examined through fitting to the real data. The production copula so obtained predicts the value added yielded by a firm with given capital and labor in a probabilistic way. 
This gives rise to paradigm shift in the economic theory for studying production activities of firms; generalization of such a canonical economic concept as the production function at a microscopic level.

To demonstrate possible applications of the production copula, we carried out a simulation for production activities of the Japanese listed firms. Also we noted that the production copula would enable us to predict the GDP with statistical uncertainty taking into account the diversity in firms' productivity.

We thus believe that the production copula will play a role of Jacob's Ladder to go back and forth between micro and macro economics. With the production copula, for instance, it is not necessary to solve such a long-standing aggregation problem as microscopic foundation of the macroscopic production function.

\section*{Acknowledgments}
We would like to thank Em.~Prof.~Masanao Aoki for encouraging
us to write this paper. Part of this research was financially supported by Hitachi Research Institute. We also appreciate the Yukawa Institute for Theoretical Physics at Kyoto University. Discussions during the YITP workshop YITP-W-07-16 on ``Econophysics III: Physical Approach to Social and Economic Phenomena" were useful to complete this paper.

%%%%%%%%%%%%%%%%%%%%%%%%%%%%%%%%%%%%%%%%%%%%%%%%%%%%%%%%%%%%%%%
\ifx\undefined\bysame
\newcommand{\bysame}{\leavevmode\hbox to\leftmargin{\hrulefill\,\,}}
\fi

\end{document}